\documentclass[pre,aps,twocolumn,showpacs]{revtex4-1}

\usepackage{graphicx}
\usepackage{amsfonts}
\usepackage{amsmath}
\usepackage{mathtools}
\usepackage{float}
\usepackage{epsfig}
\usepackage{color}
\usepackage{gensymb}
\usepackage{multirow}
\usepackage[mathscr]{euscript}
\usepackage{pifont}

\newcommand{\cmark}{\ding{51}}%
\newcommand{\xmark}{\ding{55}}%

\newcommand{\ee}[1]{\ensuremath{10^{#1}}}

\begin{document}

\title{Ground states of stealthy hyperuniform potentials: II. Stacked-slider phases}

\author{G. Zhang}


\affiliation{\emph{Department of Chemistry}, \emph{Princeton University},
Princeton, New Jersey 08544, USA}

\author{F. H. Stillinger}


\affiliation{\emph{Department of Chemistry}, \emph{Princeton University},
Princeton, New Jersey 08544, USA}

\author{S. Torquato}

\email{torquato@electron.princeton.edu}

\affiliation{\emph{Department of Chemistry, Department of Physics,
Princeton Institute for the Science and Technology of
Materials, and Program in Applied and Computational Mathematics}, \emph{Princeton University},
Princeton, New Jersey 08544, USA}


\begin{abstract}
Stealthy potentials, a family of long-range isotropic pair potentials, produce infinitely degenerate disordered ground states at high densities and crystalline ground states at low densities in $d$-dimensional
Euclidean space $\mathbb{R}^d$.  In the previous paper in this series,
we numerically studied the entropically favored ground states in the canonical ensemble in the zero-temperature limit across the first three Euclidean space dimensions. 
In this paper, we investigate using both numerical and theoretical techniques metastable stacked-slider
phases, which are part of the ground-state manifold of stealthy potentials at densities
in which crystal ground states are favored entropically.
Our numerical results enable us to devise analytical models of this phase in two, three and higher dimensions. Utilizing this model, we estimated the size of the feasible region in configuration space of the stacked-slider phase, finding it to be smaller than that of crystal structures in the infinite-system-size limit, which
is consistent with our recent previous work.
In two dimensions, we also determine exact expressions for the pair correlation function and structure factor of the analytical model of stacked-slider phases and analyze the connectedness of the ground-state manifold of stealthy potentials in this density regime.
We demonstrate that stacked-slider phases are distinguishable states of matter; they are nonperiodic, statistically  anisotropic structures
that possess long-range orientational order but have zero shear modulus.
We outline some possible future avenues of research to elucidate our understanding of this unusual phase of matter.

\end{abstract}

\maketitle

\section{Introduction}

A fundamental problem of statistical mechanics is the determination of the phase diagram of interacting many-particle systems. A substantial variety of pair interactions can produce a dramatic diversity of macroscopic phases, including crystals \cite{sands2012introduction}, quasicrystals \cite{shechtman1984metallic, levine1984quasicrystals, levine1986quasicrystals, bindi2009natural, dotera2014mosaic}, liquid crystals \cite{chandrasekhar1992liquid}, hexatic phases \cite{kosterlitz1973ordering, bernard2011two, prestipino2011hexatic,kapfer2015two}, disordered hyperuniform systems \cite{torquato2003local, donev2005unexpected, uche2006collective, torquato2008point, zachary2009hyperuniformity, jiao2014avian, torquato2015ensemble}, and liquids \cite{tabor1991gases}. While crystals and liquids are the most common condensed states of matter, there are other states in between. For example, quasicrystals and liquid crystals both have anisotropy and long-range orientational order, like crystals, but lack long-range translational order, similar to liquids. Other phases with features that lie between crystals and liquids include disordered hyperuniform systems, which are disordered but behave more like crystals in the way in which they suppress long-range density fluctuations \cite{torquato2003local, jiao2014avian}. 

A family of long-range isotropic pair potentials, called stealthy potentials, produces infinitely degenerate disordered hyperuniform classical ground states at high densities in $d$-dimensional
Euclidean space $\mathbb{R}^d$ \cite{uche2006collective, torquato2015ensemble, uche2004constraints, batten2008classical, batten2009interactions, batten2011inherent, zhang2015ground}.
Stealthy potentials are often specially constructed such that finding a ground state is equivalent to constraining the structure factor $S(\mathbf k)$ to be zero for all wave vectors $\mathbf k$ such that $0<|\mathbf k| \le K$, where $K$ is some radial cutoff value.
A dimensionless measure of the relative fraction of constrained degrees of freedom (proportional to $K^d$) compared to the total number of degrees of freedom, $\chi$, controls the degree of order and degeneracy of the ground states of these potentials.

In the preceding paper \cite{zhang2015ground},
we numerically studied the entropically favored ground states, i.e., configurations most likely to appear in the canonical ensemble in the zero-temperature limit, of stealthy potentials. 
We found that entropically favored ground states are disordered for $\chi<1/2$, and crystalline for $\chi>1/2$ up to a certain critical value \cite{torquato2015ensemble}.

The main focus of this paper is the investigation of stacked-slider phases, which are metastable states that are part of the ground-state manifold for some $\chi$ above 1/2, although not entropically favored. Stacked-slider phases were first discovered in two dimensions in Ref.~\onlinecite{uche2004constraints} and were originally called wavy crystals because they were observed to consist of particle columns that display a meandering displacement away from linearity. However, we will see that ``stacked-slider phases'' for arbitrary dimensions is a more suitable name for this phase and this designation will be used henceforth.

The authors of Ref.~\onlinecite{uche2004constraints} easily distinguished stacked-slider phases from crystal phases by a lack of periodicity in direct space and a lack of Bragg peaks in its diffraction pattern.
Distinguishing stacked-slider phases and disordered phases, on the other hand, was based on a different property.
In disordered phases, all $\mathbf k$'s such that $|\mathbf k| > K$ have positive structure factors. However, in stacked-slider phases, the structure factor at some $\mathbf k$'s such that $|\mathbf k| > K$ are implicitly constrained to vanish identically \cite{uche2004constraints}, i.e., they are induced to be zero by the constraints inside the radius $K$. The existence of implicit constraints was used to distinguish stacked-slider phases from disordered phases in Ref.~\onlinecite{uche2004constraints}.

There are still many outstanding questions concerning
stacked-slider phases. 
Can a theoretical model of stacked phases in the thermodynamic limit be devised to elucidate 
previous numerical studies? 
One disadvantage of numerical studies is that finite-size effects make it difficult to
conclude anything definitive
about the large system limit.
For example, are stacked-slider phases isotropic or anisotropic in this limit? Moreover, to what extent does the choice
of the simulation box shape affect the results? 
Were any important features of stacked-slider phases overlooked by studying finite-precision simulation results? 
Finally, because Ref.~\onlinecite{uche2004constraints} studied two dimensions only, we do not know whether stacked-slider phases exist in other dimensions.
This paper provides additional insights into these unanswered questions. 

The rest of the paper is organized as follows. In Sec.~\ref{numerical} we perform numerical studies with much higher precision than previously. The numerical results enabled us to find an analytical model of two-dimensional stacked-slider phases, presented in Sec.~\ref{model2}. We generalize this model to higher dimensions in Sec.~\ref{model}.
We demonstrate that stacked-slider phases are distinguishable states of matter; they are nonperiodic, statistically  anisotropic structures
that possess long-range orientational order but have zero shear modulus.
The model also shows that implicit constraints exist.
In Sec.~\ref{region} we use this analytical model to show that stacked-slider phases are not entropically favored in the zero-temperature limit of the canonical ensemble.
In Sec.~\ref{stability} we postulate that the transition between stacked-slider phases and disordered phases occurs at a slightly lower $\chi$ than that reported in Ref.~\onlinecite{uche2004constraints} from energy minimizations from high-temperature limit (Poisson) initial configurations.
In Sec.~\ref{conclusion} we make concluding
remarks and draw comparisons to other
common phases of matter.

\section{Numerical study of 2D stacked-slider phases}
\label{numerical}

In this section we numerically study the ground states of a stealthy potential at a variety of $\chi$'s (or densities) in two dimensions. We begin with the mathematical relations and simulation procedure in Sec.~\ref{detail}, and then present our results in Sec.~\ref{NumericalResults}. These results will suggest an analytical model of two-dimensional stacked-slider phases in Sec.~\ref{model2}.

\subsection{Mathematical relations and simulation procedure}
\label{detail}
As detailed in the preceding paper \cite{zhang2015ground} and other references \cite{torquato2015ensemble, uche2004constraints, uche2006collective, batten2008classical, batten2009interactions, batten2011inherent}, we simulate systems consisting of $N$ point particles, located at $\mathbf r^N \equiv \mathbf r_1$, $\mathbf r_2$, ..., $\mathbf r_N$, in a simulation box in $\mathbb{R}^d$ under periodic boundary conditions. The number density is $\rho=N/v_F$, where $v_F$ is the volume of the simulation box.
The particles interact with a pairwise additive potential $v(\mathbf r)$ such that its Fourier transform is:
\begin{equation}
{\tilde v}({\mathbf k})=
\begin{cases}
\displaystyle V(|\mathbf k|)& \text{if $|\mathbf k| \le K$} \\
0& \text{otherwise,}
\end{cases}
\label{stealthy}
\end{equation}
where ${\tilde v}({\mathbf k})= \int_{v_F} v(\mathbf r) \exp(-i \mathbf k \cdot \mathbf r) d\mathbf r$ is the Fourier transform of the pair potential $v(\mathbf r)$, $V(k)$ is a positive function, and $K$ is a constant.

Under such potential, the total potential energy of the system can be calculated in the Fourier space
\begin{equation}
\Phi({\bf r}^N) =\frac{1}{2v_F} \sum_{\bf k} V(|\mathbf k|)|{\tilde n}({\bf k})|^2 +\Phi_0,
\label{pot_f2}
\end{equation}
where the sum is over all reciprocal lattice vector $\mathbf k$'s of the simulation box such that $0<|\mathbf k|\le K$, ${\tilde n}({\bf k})= \sum_{j=1}^{N} \exp(-i{\bf k \cdot r}_j)$, and
\begin{equation}
\Phi_0=[N(N-1) - N \sum_{\bf k} {\tilde v}({\bf k})]/2v_F
\end{equation}
is a constant independent of the particle positions $\mathbf r^N$. Thus, the first term on the right-hand side of Eq.~\eqref{pot_f2} is the only configuration-dependent contribution to the potential energy
\begin{equation}
\Phi^* ({\bf r}^N) =\frac{1}{2v_F} \sum_{0<|\mathbf k| \le K} V({\bf k})|{\tilde n}({\bf k})|^2.
\label{pot_f3}
\end{equation}
Since $V(k)>0$ and $v_F>0$, Eq.~\eqref{pot_f3} shows that $\Phi^* ({\bf r}^N)\ge 0$. Therefore, if configurations such that $\Phi^* ({\bf r}^N)=0$ exist, then they are the classical ground states of this potential. These configurations are achieved by constraining ${\tilde n}({\mathbf k})$ to zero for all $0<|\mathbf k| <K$ and are said to be stealthy up to $K$.
Since ${\tilde n}({\mathbf k})$ is related to the structure factor $S(\mathbf k)$ by $S(\mathbf k)=|{\tilde n}({\mathbf k})|^2/N$ for every $\mathbf k \neq \mathbf 0$, constraining ${\tilde n}({\mathbf k})$ to zero is equivalent to constraining $S(\mathbf k)$ to zero.
Let $M$ be half the number of $\mathbf k$ points in the summation of Eq.~\eqref{pot_f2} \cite{independentconstraints}; the parameter
\begin{equation}
\chi=\frac{M}{d(N-1)}
\label{chi}
\end{equation}
determines the degree to which the ground states are constrained and therefore the degeneracy and disorder of the ground states \cite{uche2004constraints}.
For a fixed $K$, the parameter $\chi$ is inversely proportional to the density \cite{batten2009interactions, torquato2015ensemble, zhang2015ground}. When $\chi \le \chi_{max}^*$, where $\chi_{max}^*$ is a dimension-dependent constant, all the constraints are indeed satisfiable, thus $\Phi^* ({\bf r}^N)$ of the classical ground states is zero \cite{torquato2015ensemble}.
The $\chi$ values we study in this paper are always less than $\chi_{max}^*$.

In this section we choose $N=100$, $K=1$, and $V(k)=1$. The relatively small choice of $N$ increases the precision of the ground states we find. We will see that high precision is important in extracting an analytical model from numerical results. The constant $K$ and the magnitude of $V(k)$ simply set the length scale and the energy scale. Although the function form of $V(k)$ could theoretically affect the probability of sampling different parts of the ground-state manifold, it does not affect the manifold itself \cite{torquato2015ensemble, zhang2015ground}. As explained in Ref.~\onlinecite{zhang2015ground}, we use a rhombic simulation box with a $60^\circ$ interior angle to alleviate finite-size effect.

The ground states reported in this section are produced by the following steps.
\begin{enumerate}
\setlength\itemsep{0em}
\item Start from a Poisson (i.e., ideal gas) initial configuration.
\item Minimize $\Phi^* ({\bf r}^N)$ [in Eq.~\eqref{pot_f3}] using the low-storage BFGS algorithm \cite{nocedal1980updating, liu1989limited, nlopt}.
\item Minimize $\Phi^* ({\bf r}^N)$ using the MINOP algorithm \cite{dennis1979two}.
\item If $\Phi^* ({\bf r}^N)< \ee{-20}$, we successfully find a relatively high-precision ground state.
\item Otherwise, what we find is either an imprecise ground state or a local minimum of $\Phi^* ({\bf r}^N)$. Therefore, we discard this configuration.
\end{enumerate}
As detailed in Ref.~\onlinecite{zhang2015ground}, the low-storage BFGS algorithm is the fastest in minimizing $\Phi^* ({\bf r}^N)$, while the MINOP algorithm finds the most precise ground states. Therefore, we minimize $\Phi^* ({\bf r}^N)$ using these two algorithms consecutively to maximize both efficiency and precision.

These steps are performed $N_t$ times for a variety of simulation box side lengths (and therefore a variety of $\chi$'s), listed in Table~\ref{Nt} \cite{successrate}.
As detailed in Ref.~\onlinecite{uche2004constraints}, for a finite system, only certain values of $\chi$ are allowed. The $\chi$ values in Table~\ref{Nt} contain all possible choices in the range $0.5<\chi<\chi_{max}^*$, which covers the previously reported stacked-slider phase regime in two dimensions, $0.57... \le \chi < 0.77...$ \cite{uche2004constraints, batten2009interactions}.
Except for $\chi=0.8787\ldots$, where we could not precisely identify ground states, we plot the real-space configuration and reciprocal-space structure factor of at least 50 successful energy minimized results and visually inspect them. We divide them into different categories based on their appearances and then present representative configurations below.

\begin{table}
\setlength{\tabcolsep}{8pt}
\caption{The $\chi$ values, number of trials $N_t$, and number of successes $N_s$ for each simulation box side length $L$.}
\begin{tabular}{c c c c }
\hline
$L$ & $\chi$ & $N_t$ & $N_s$\\
\hline
56 & 0.5303\ldots & 1000000 & 15615\\
57 & 0.5606\ldots & 199915 & 17127\\
58 & 0.5909\ldots & 200000 & 411\\
59 & 0.6060\ldots & 199965 & 8875\\
60 & 0.6363\ldots & 1000000 & 27788\\
62 & 0.6666\ldots & 1000000 & 76727\\
63 & 0.6818\ldots & 1000000 & 157501\\
64 & 0.7121\ldots & 200000 & 119563\\
65 & 0.7424\ldots & 1000000 & 165203\\
66 & 0.7575\ldots & 200000 & 80258\\
68 & 0.7878\ldots & 200000 & 2577\\
70 & 0.8787\ldots & 200000 & 0\\
\hline
\end{tabular}
\label{Nt}
\end{table}

\subsection{Results}
\label{NumericalResults}

\begin{figure*}
\begin{center}
\includegraphics[width=0.24\textwidth, trim=100 200 100 150]{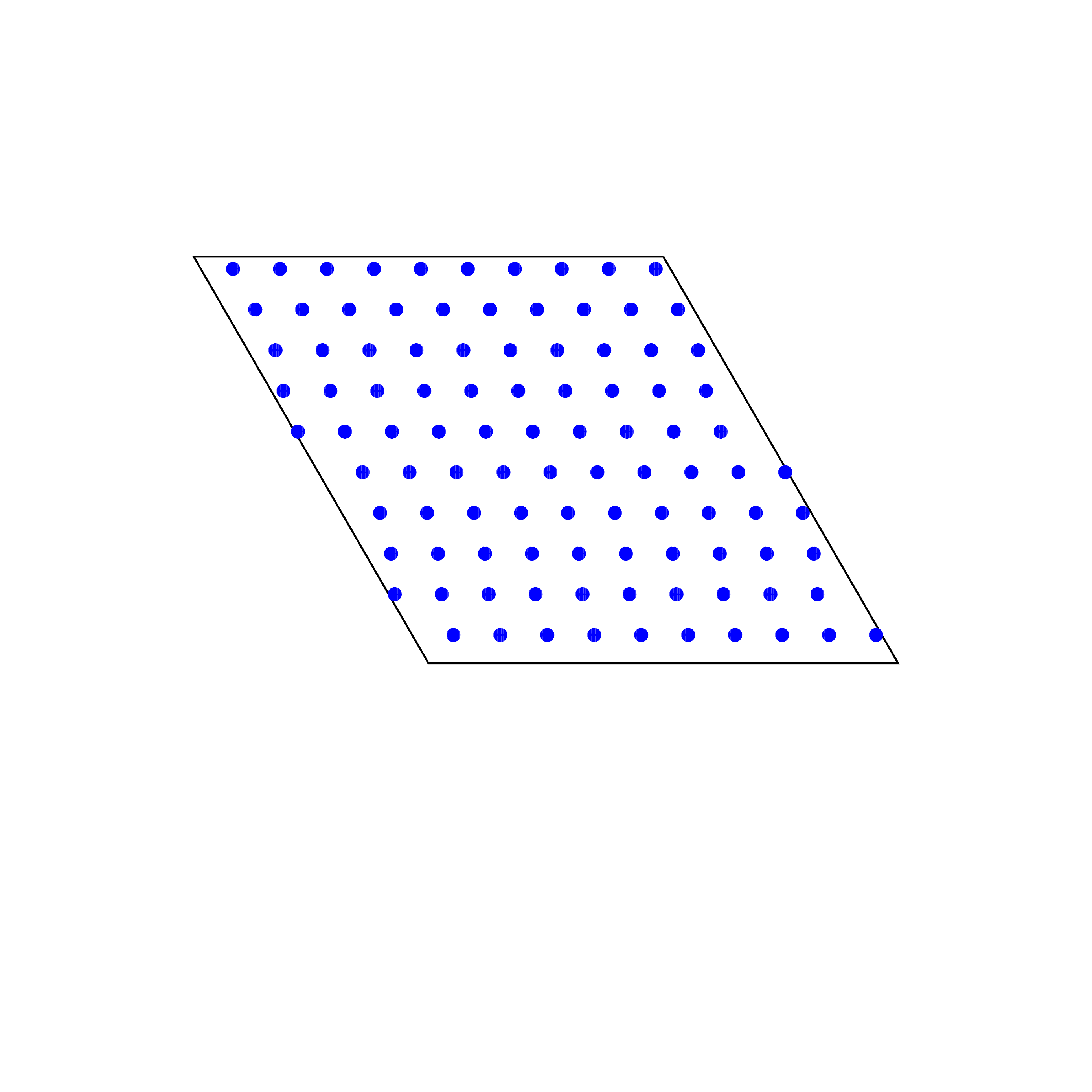}
\includegraphics[width=0.31\textwidth, trim=0 0 0 50]{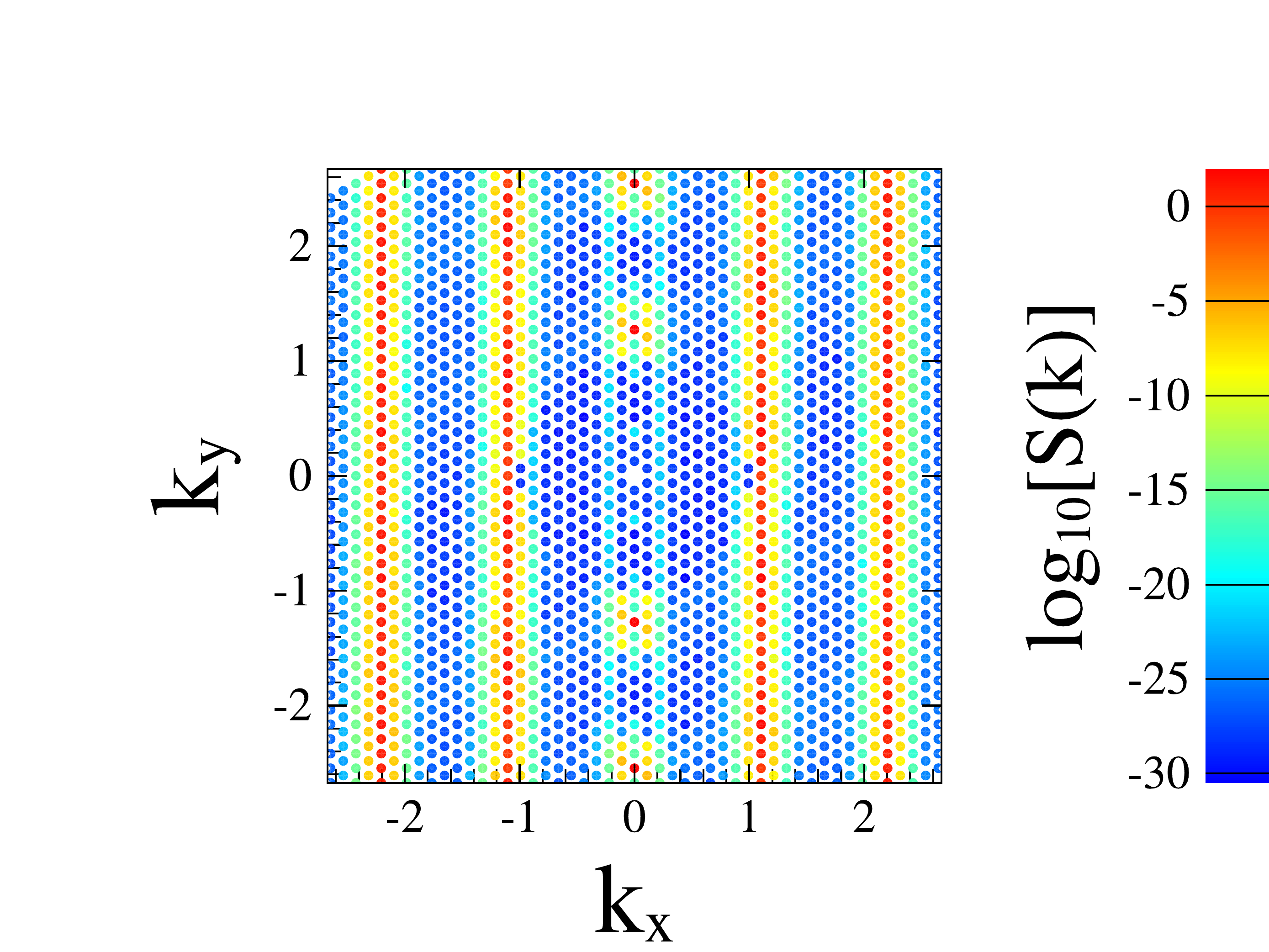}

\includegraphics[width=0.24\textwidth, trim=100 200 100 150]{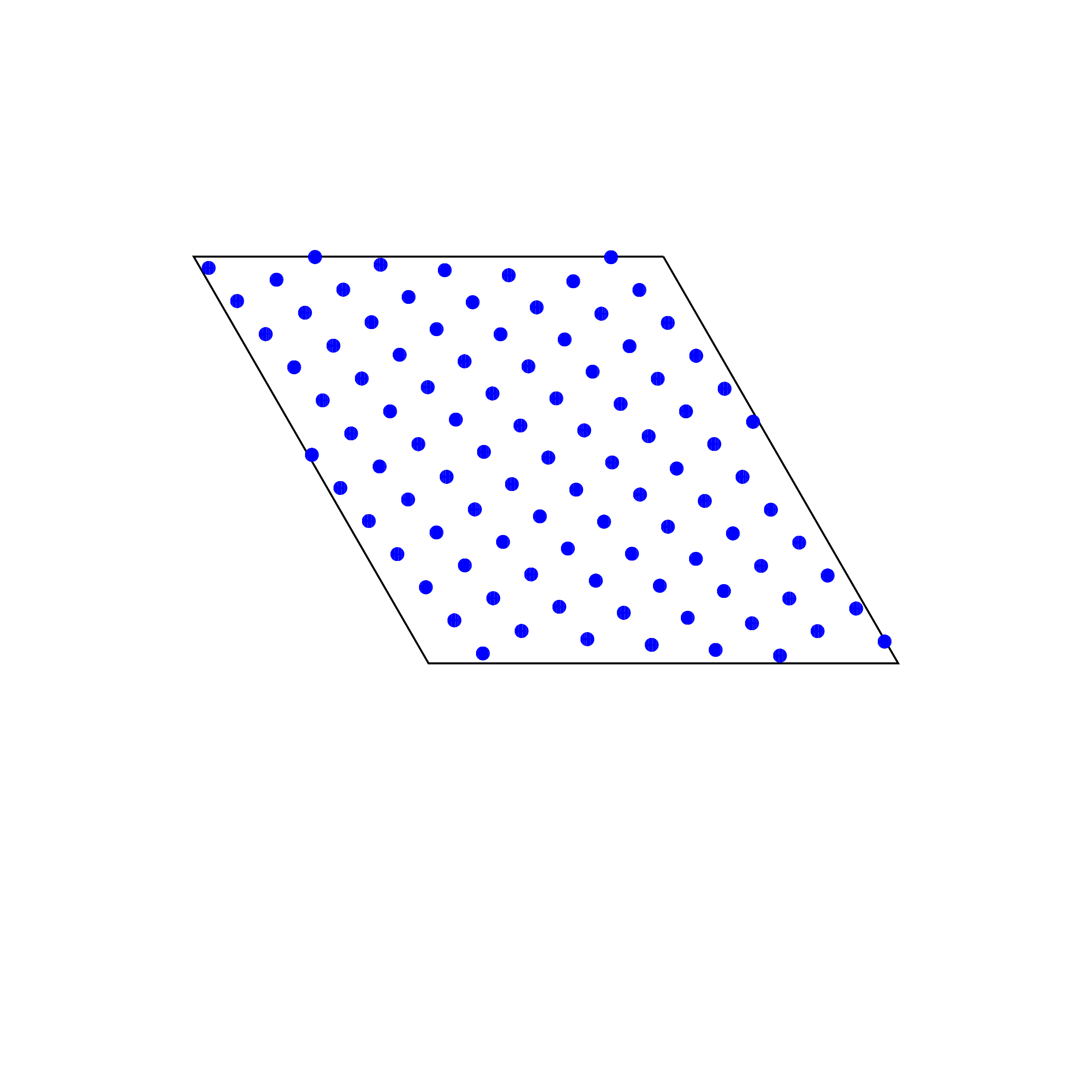}
\includegraphics[width=0.31\textwidth, trim=0 0 0 50]{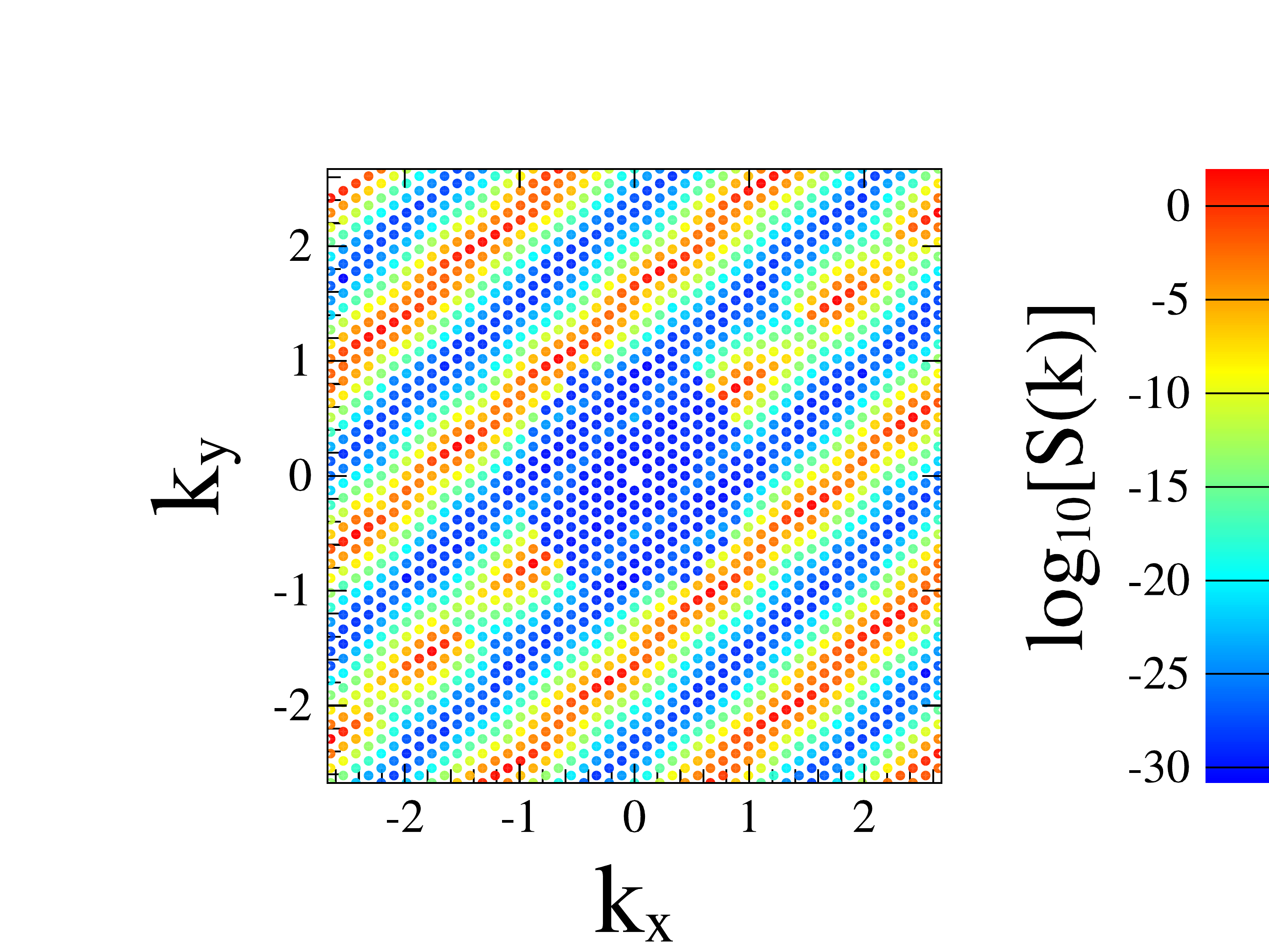}

\includegraphics[width=0.24\textwidth, trim=100 200 100 150]{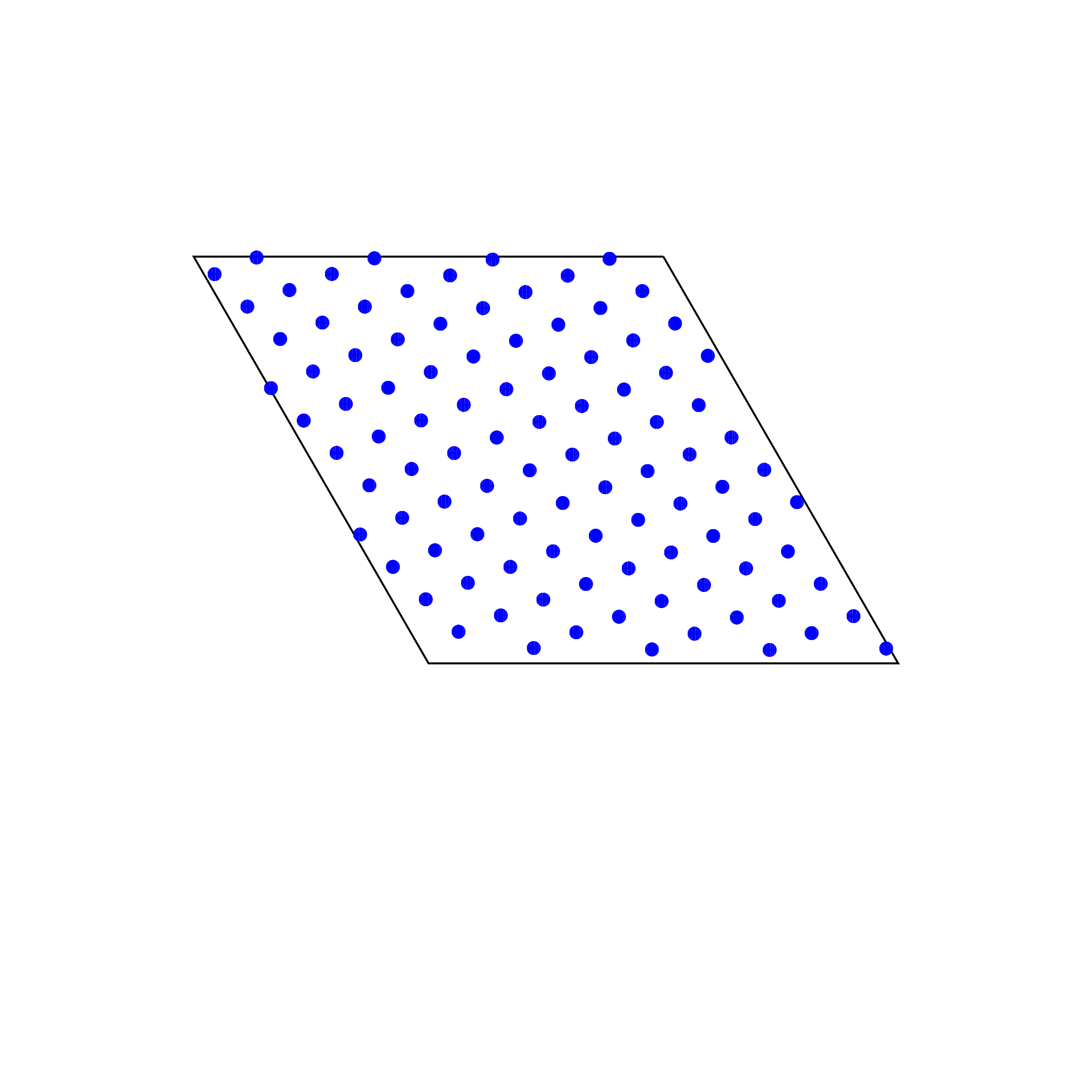}
\includegraphics[width=0.31\textwidth, trim=0 0 0 50]{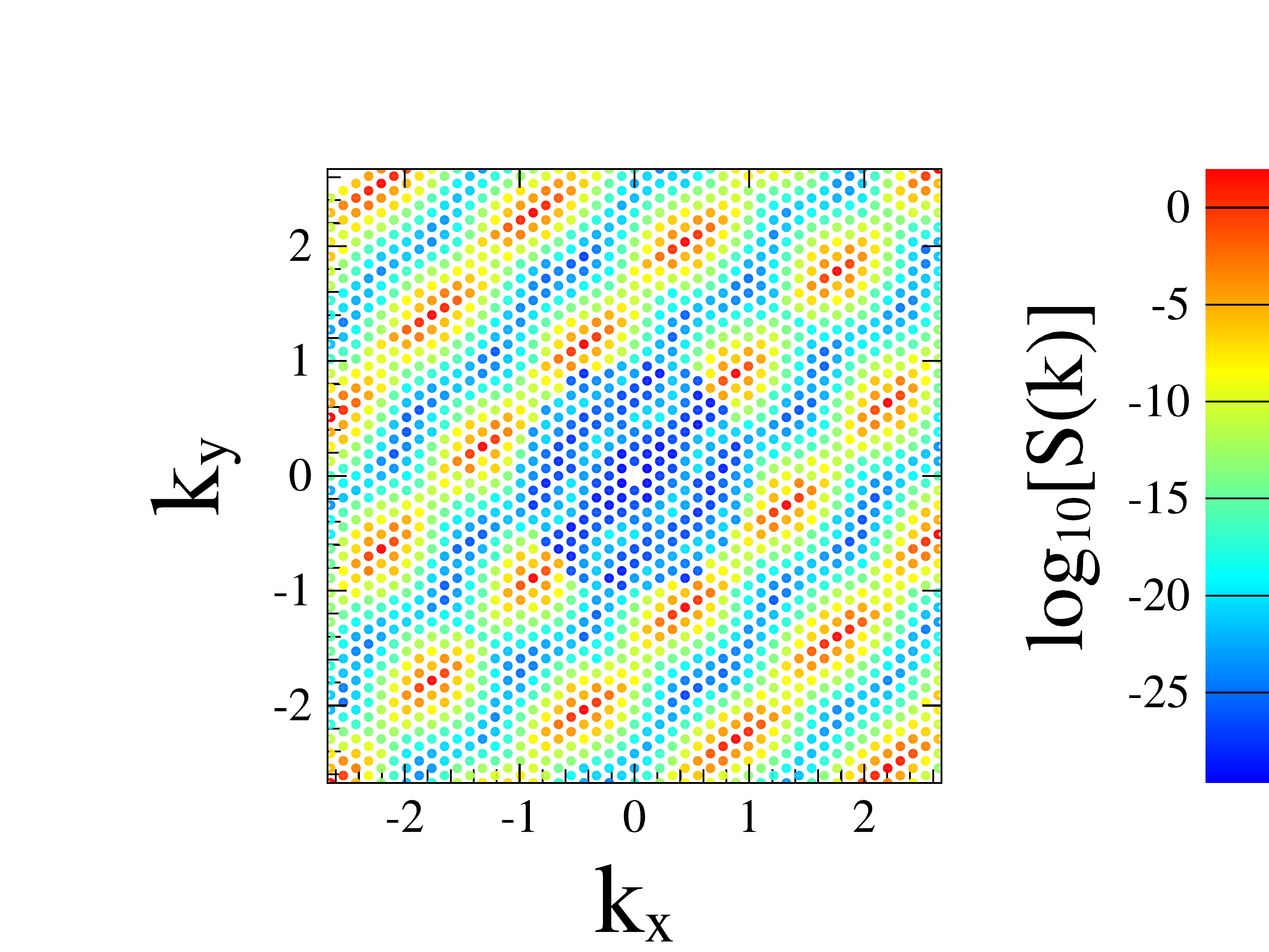}

\includegraphics[width=0.24\textwidth, trim=100 200 100 150]{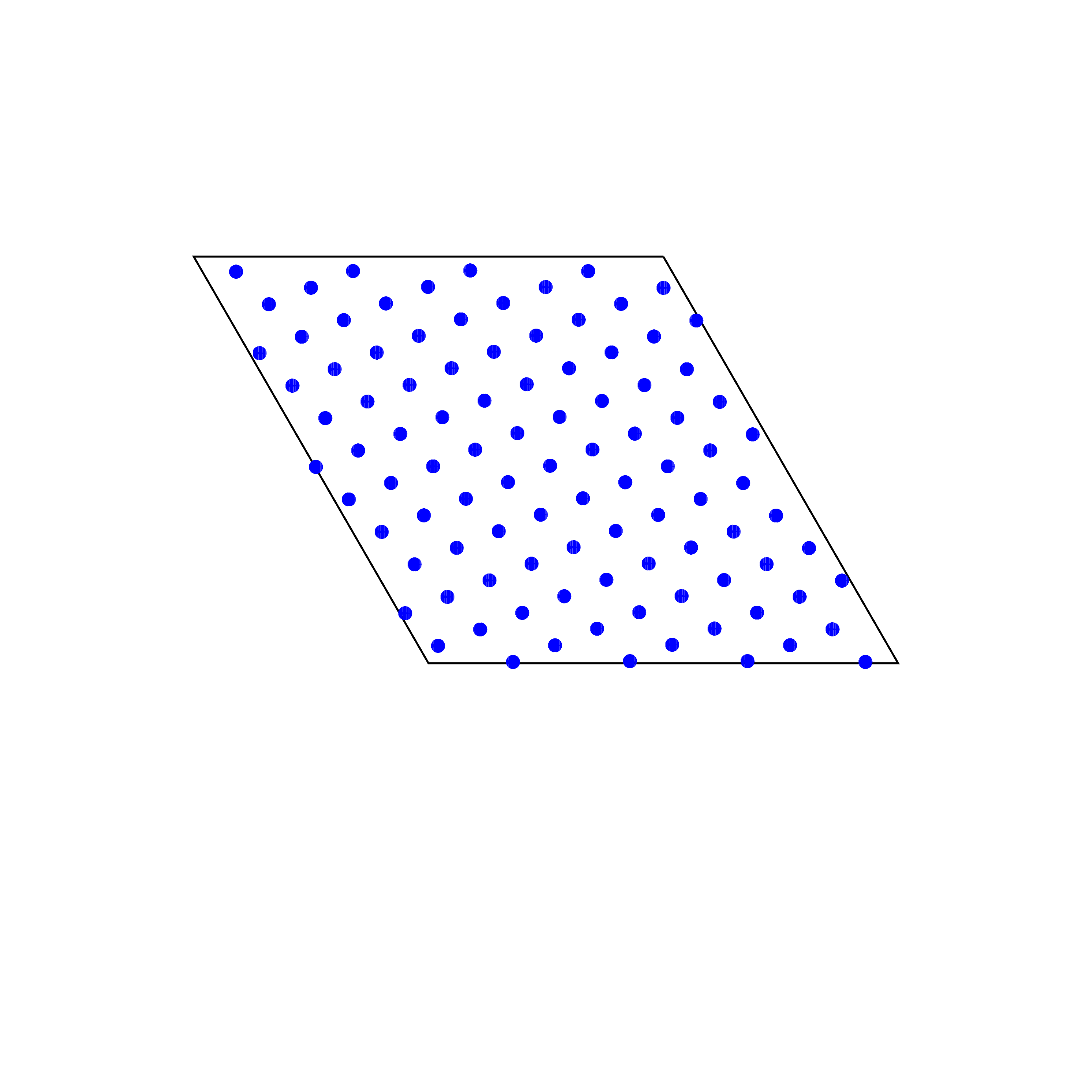}
\includegraphics[width=0.31\textwidth, trim=0 0 0 50]{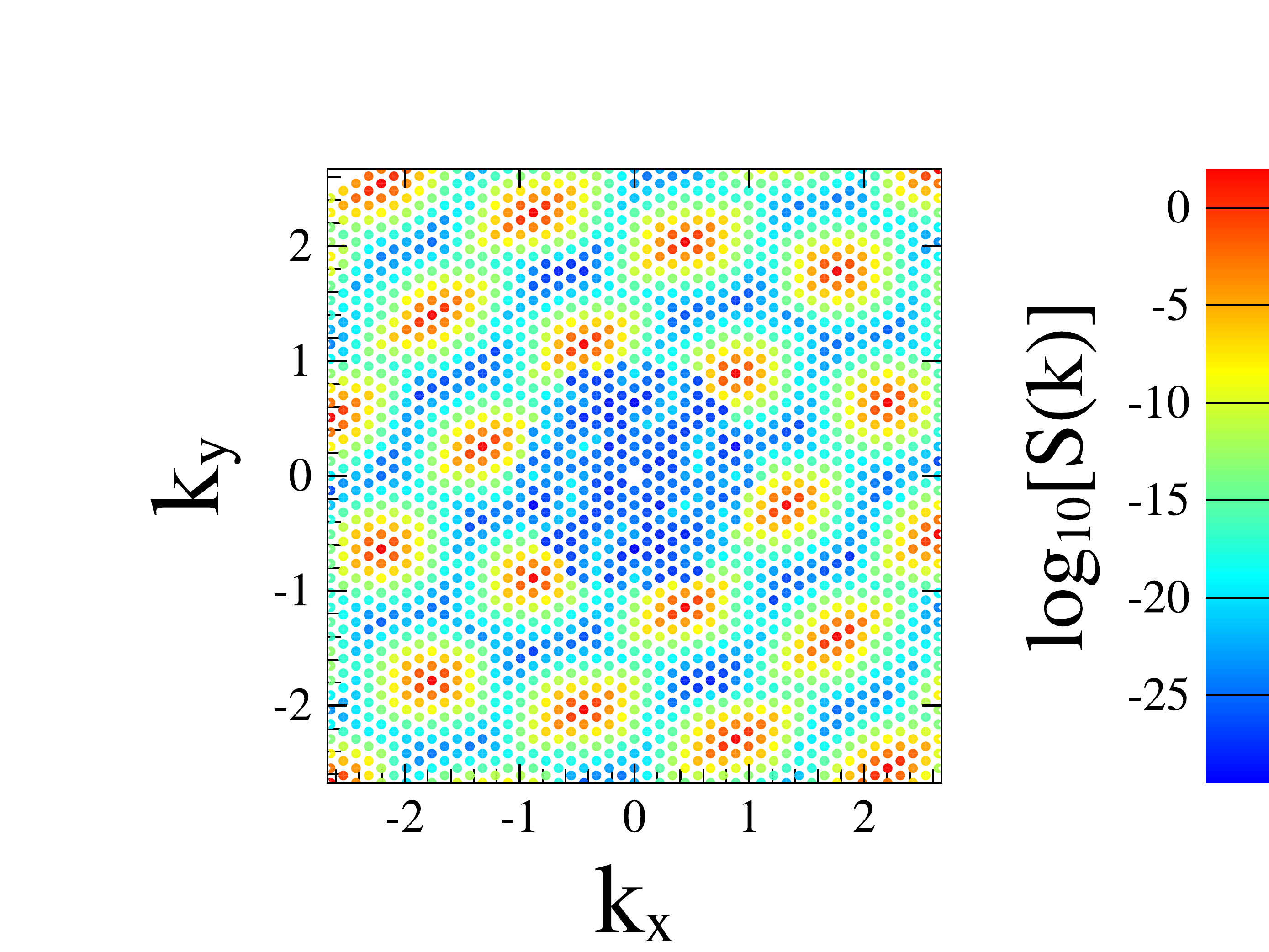}

\end{center}
\caption{(Color online) Four representative numerically obtained ground-state configurations at $\chi=0.5606\ldots$ (left) and their corresponding structure factors (right), where colors indicate intensity values at reciprocal lattice points.
}
\label{fig_lowchi}
\end{figure*}

\begin{figure*}
\begin{center}
\includegraphics[width=0.24\textwidth, trim=100 200 100 150]{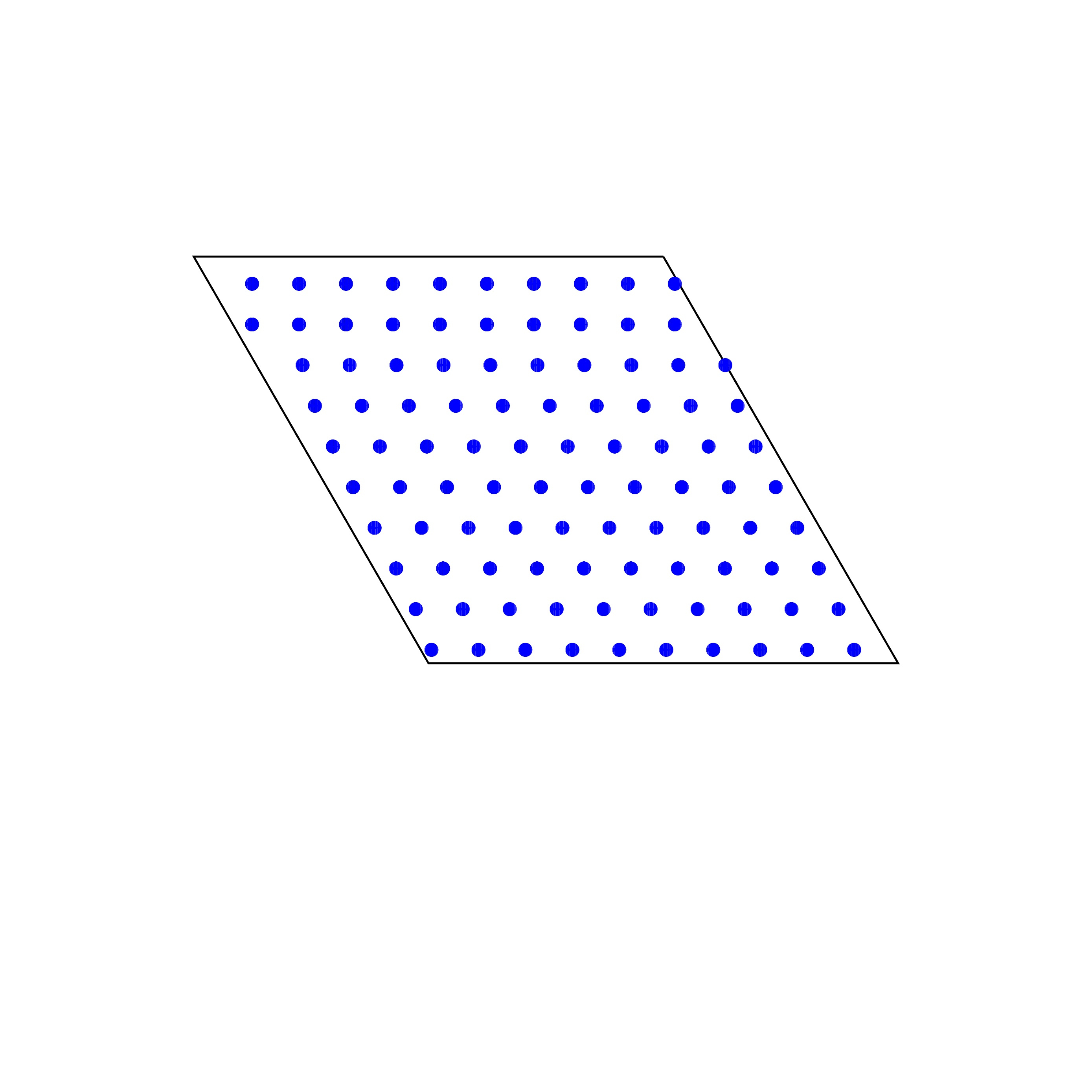}
\includegraphics[width=0.31\textwidth, trim=0 0 0 50]{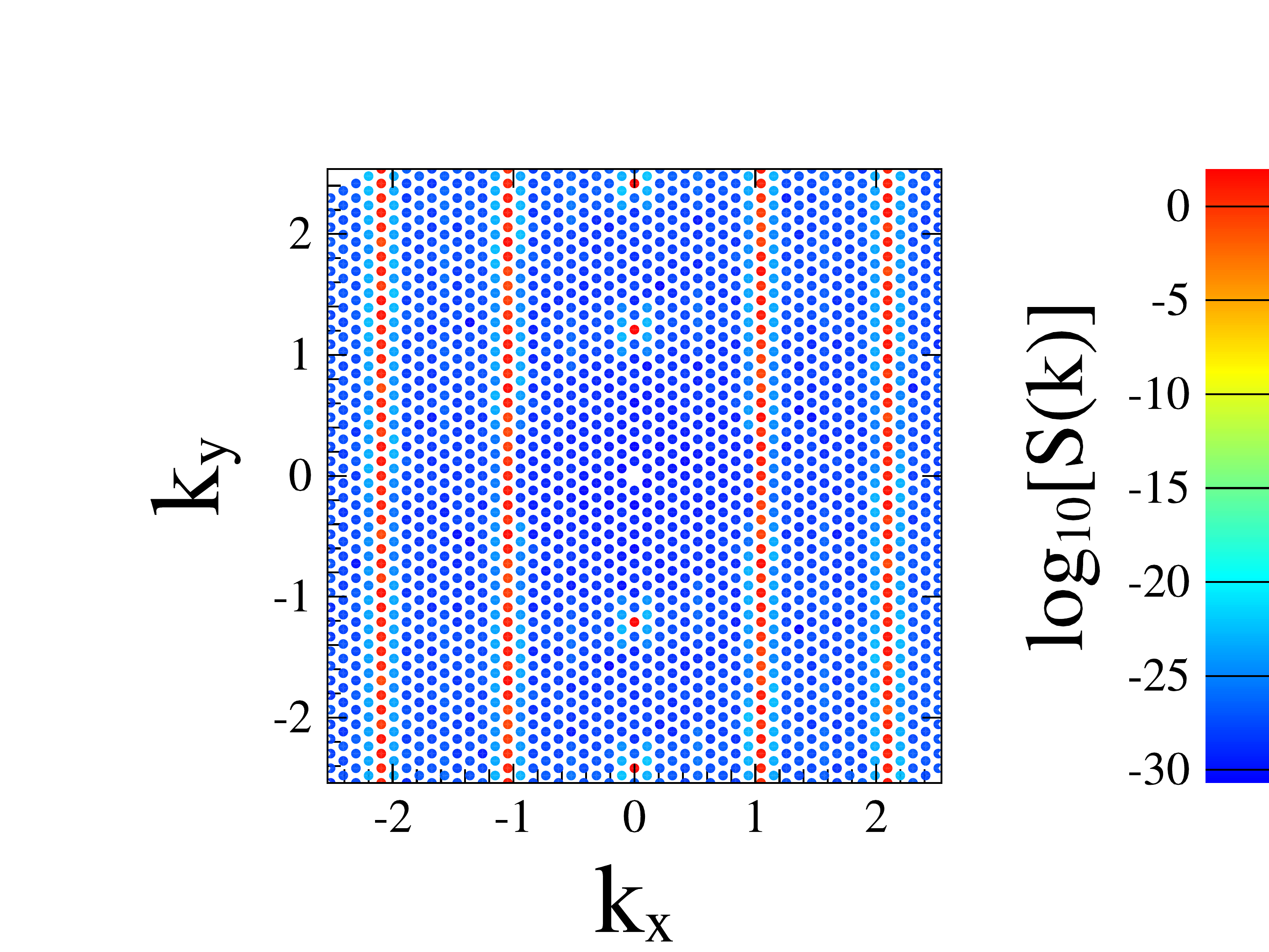}

\end{center}
\caption{(Color online) A numerically obtained ground-state configuration at $\chi=0.6363\ldots$ (left) and the corresponding structure factor (right), where colors indicate intensity values at reciprocal lattice points.}
\label{fig_medchi_tri}
\end{figure*}

\begin{figure*}
\begin{center}
\includegraphics[width=0.25\textwidth, trim=100 200 100 150]{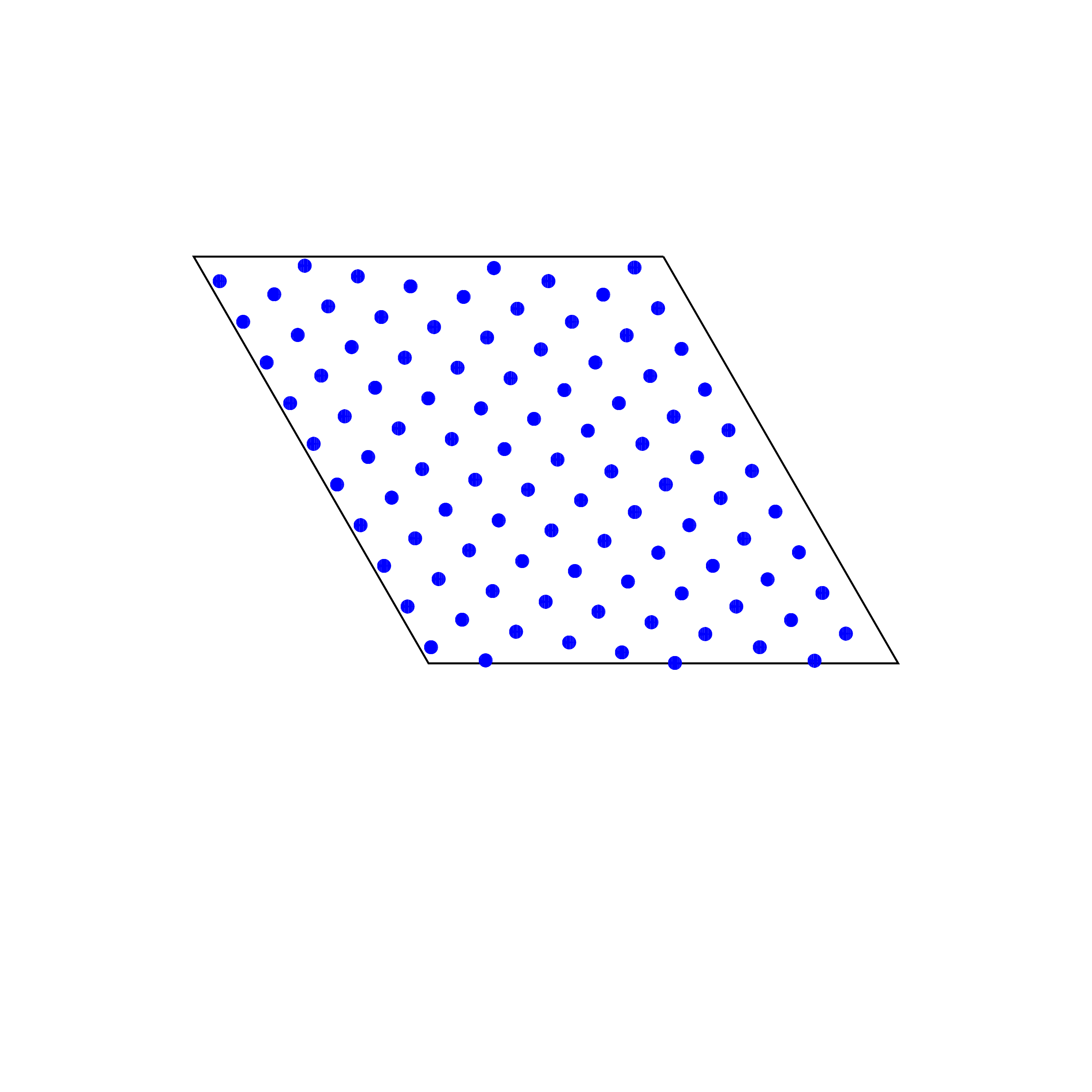}
\includegraphics[width=0.33\textwidth, trim=0 0 0 50]{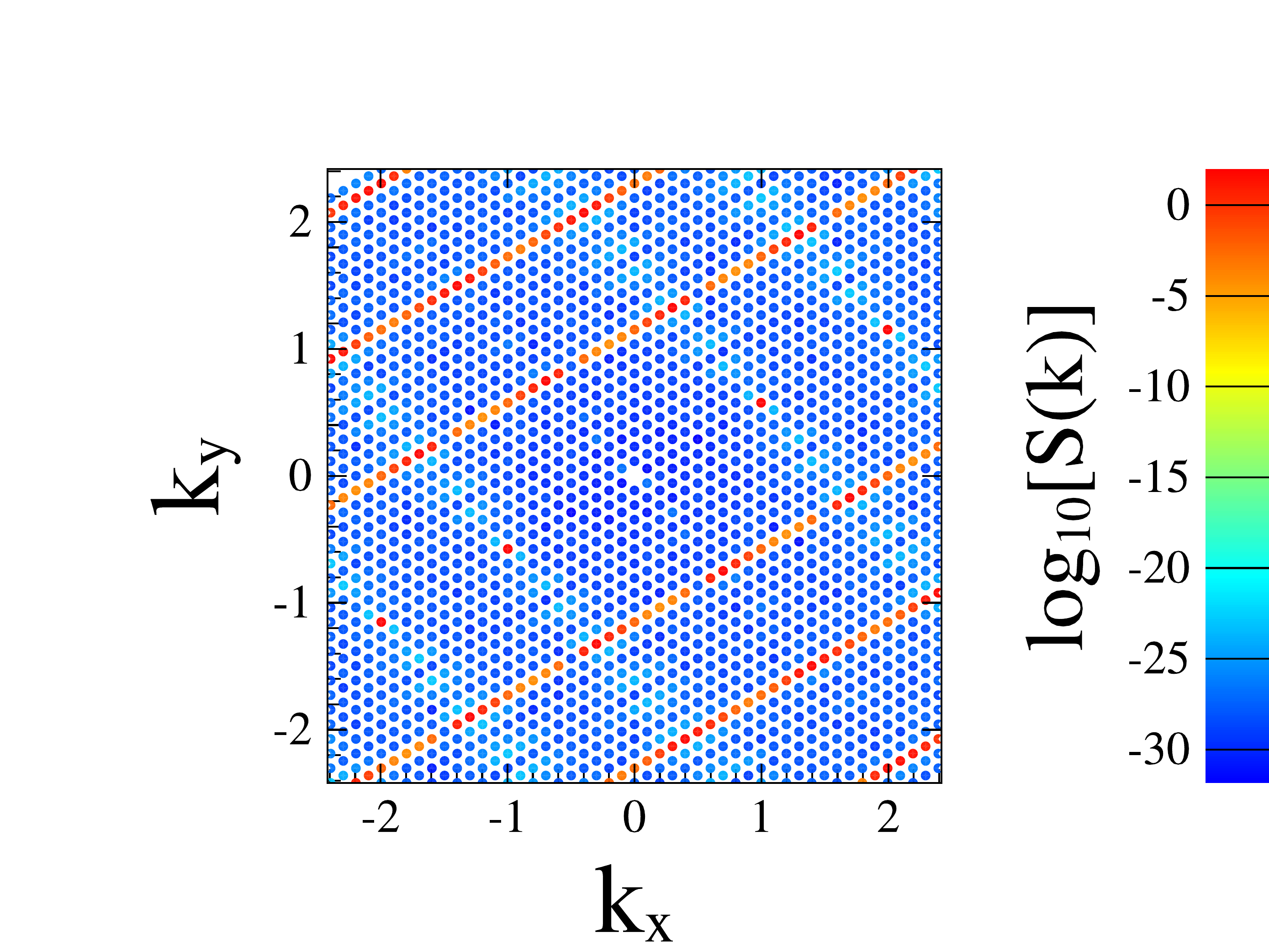}

\includegraphics[width=0.25\textwidth, trim=100 200 100 150]{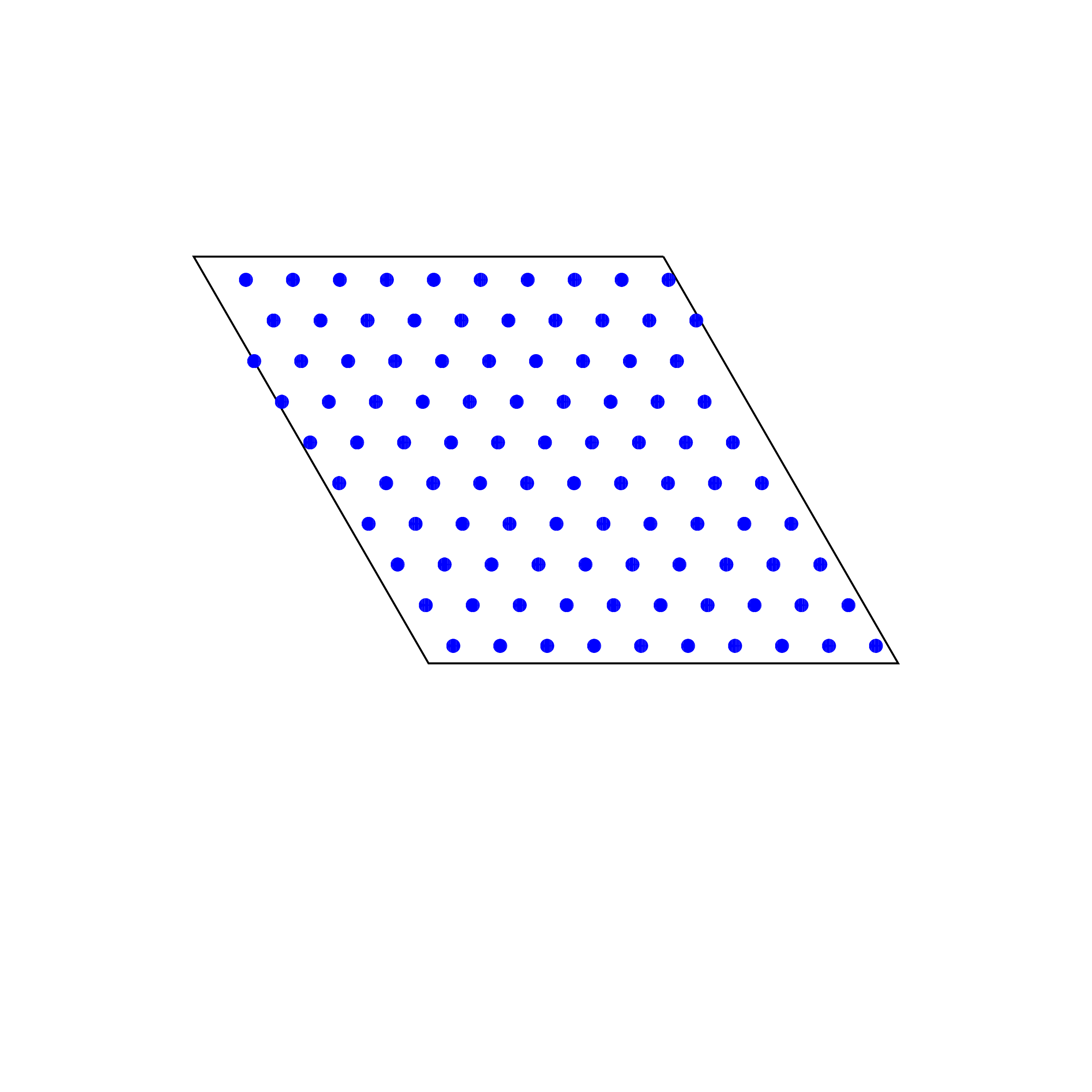}
\includegraphics[width=0.33\textwidth, trim=0 0 0 50]{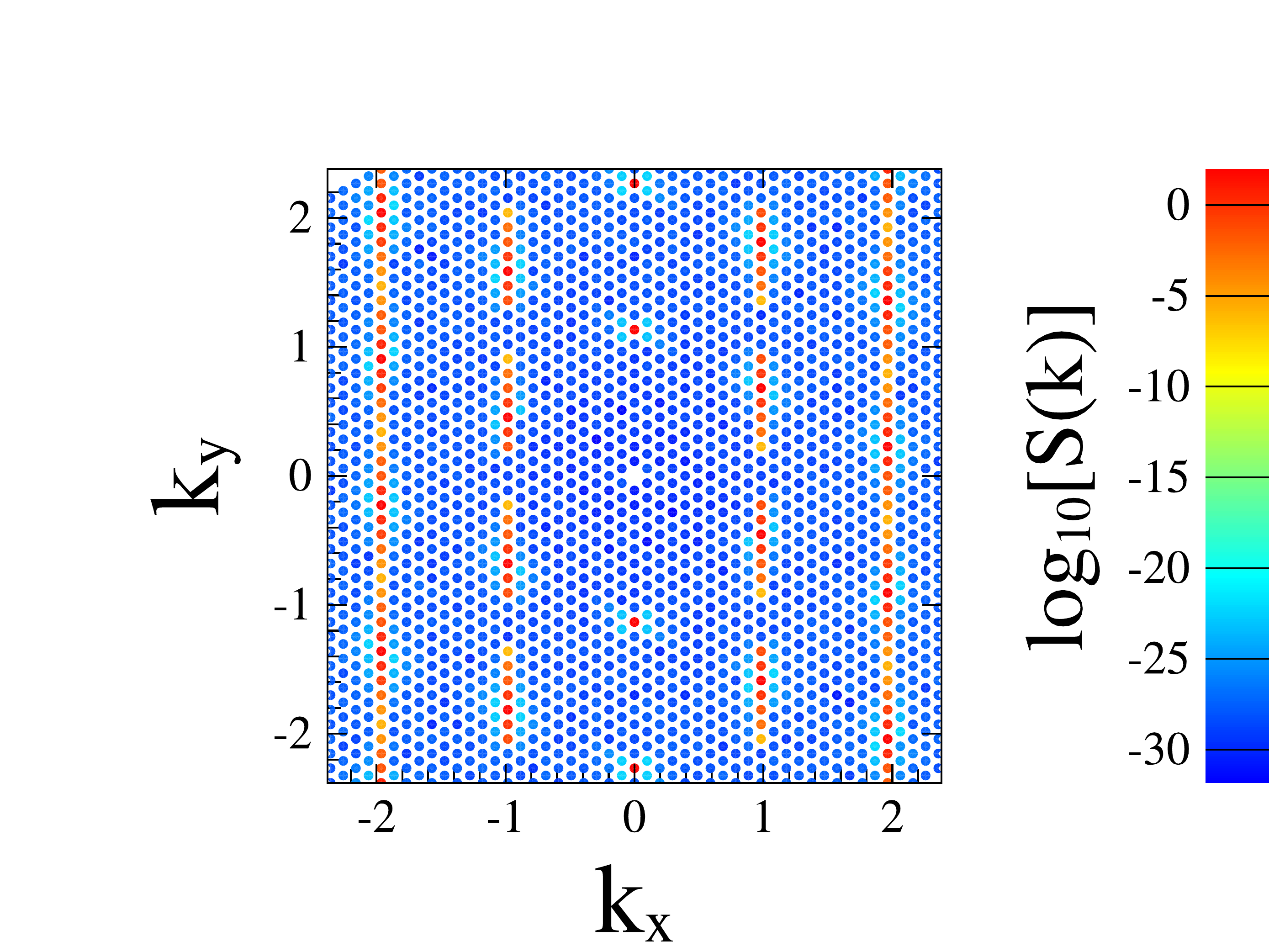}

\includegraphics[width=0.25\textwidth, trim=100 200 100 150]{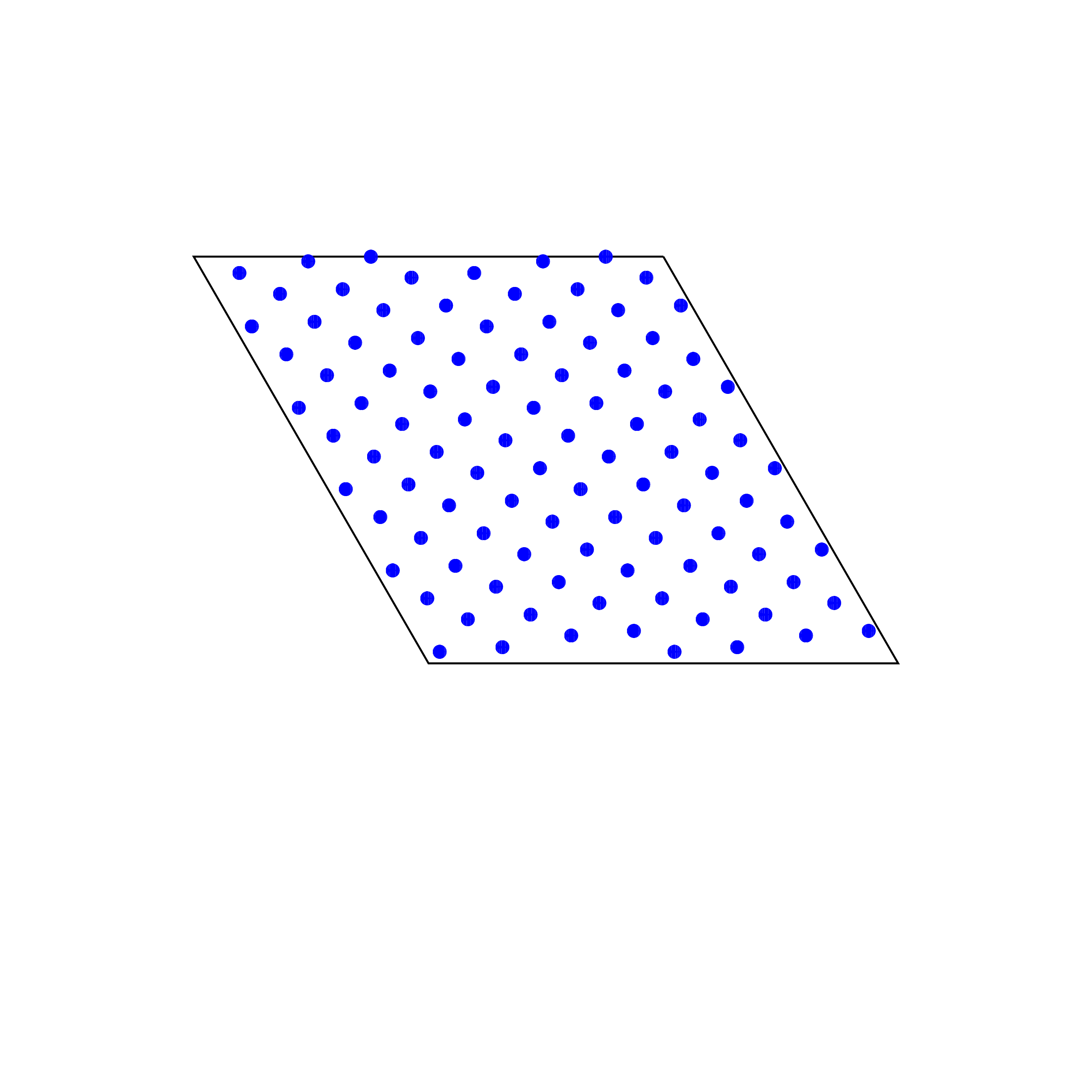}
\includegraphics[width=0.33\textwidth, trim=0 0 0 50]{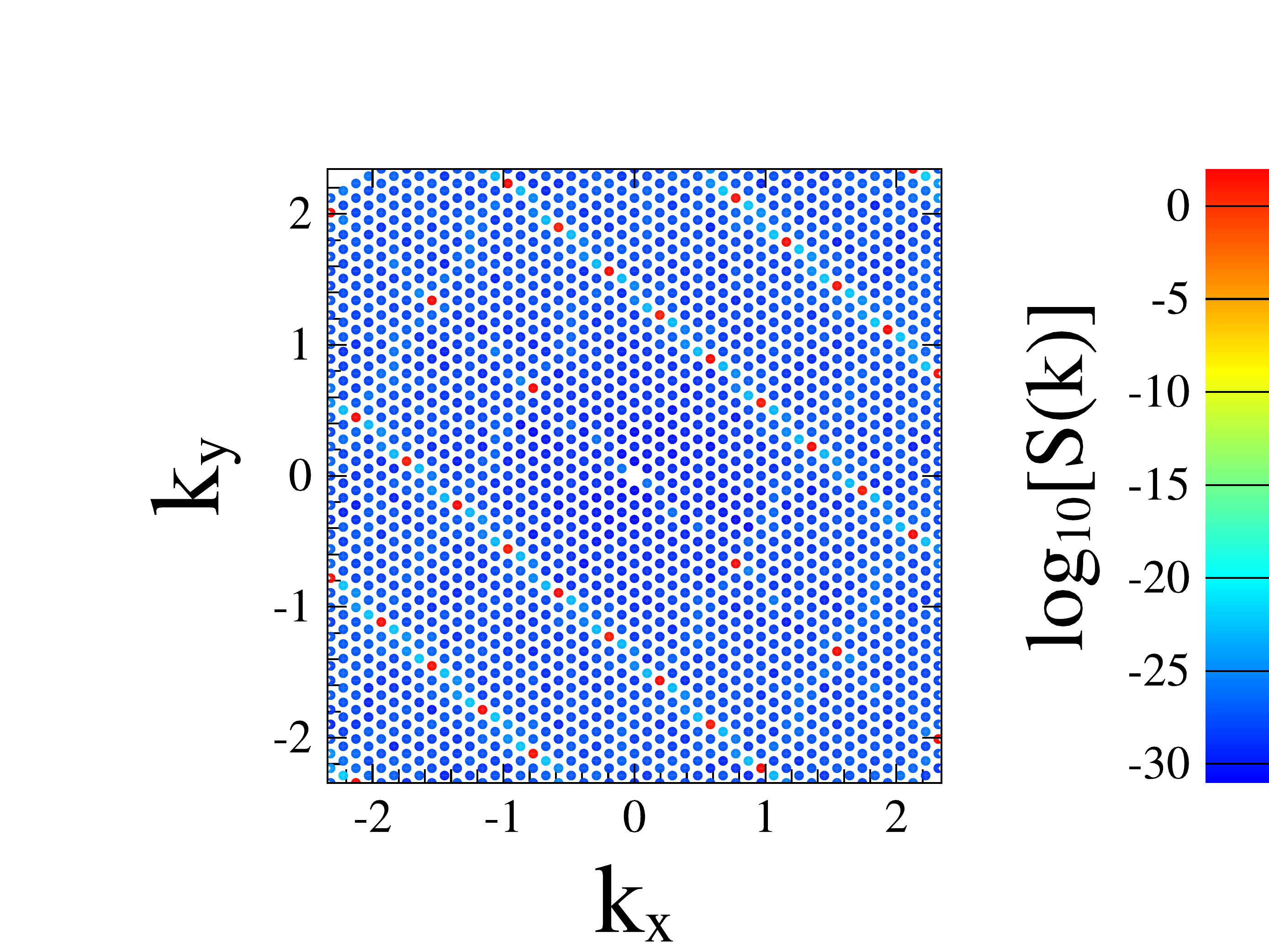}

\includegraphics[width=0.25\textwidth, trim=100 200 100 150]{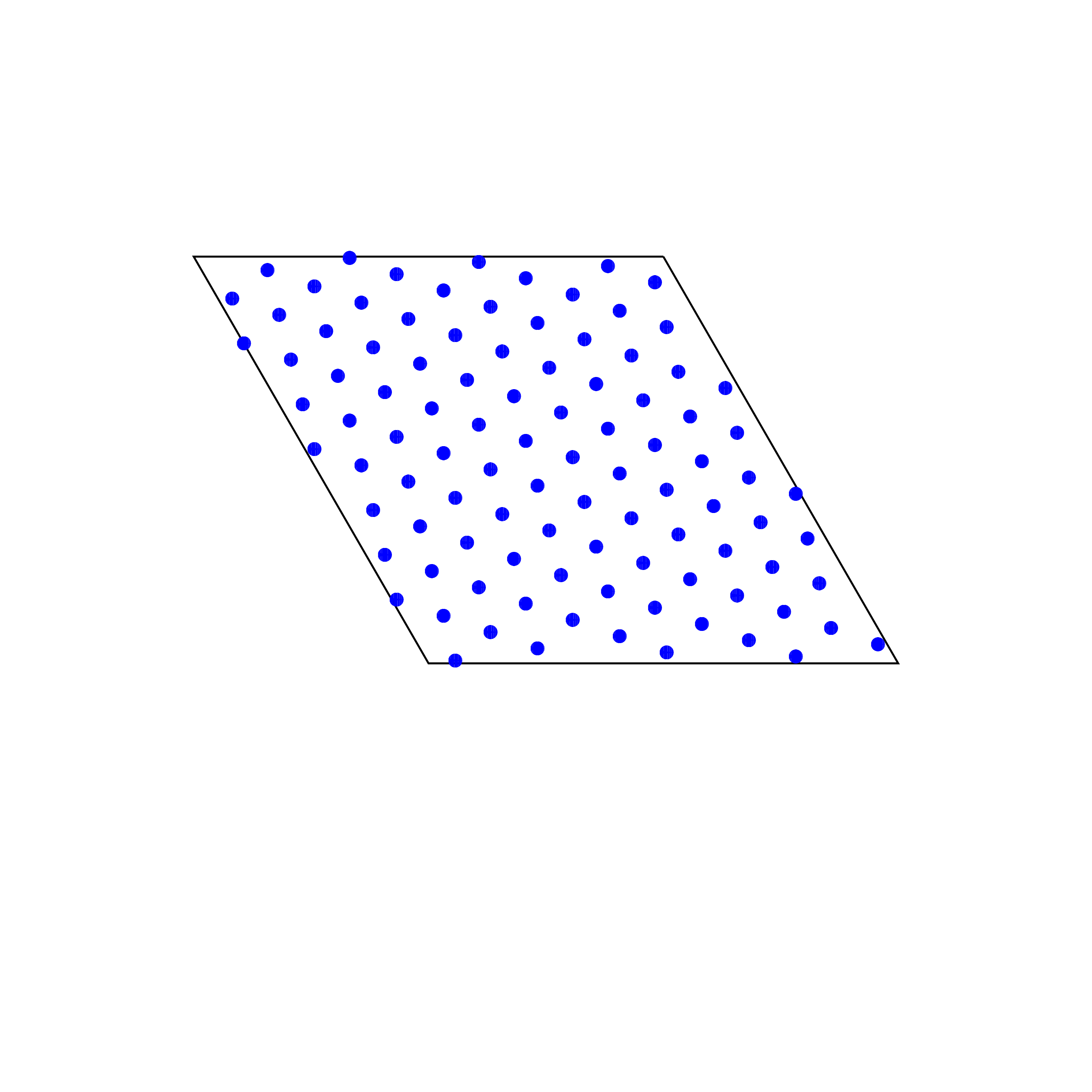}
\includegraphics[width=0.33\textwidth, trim=0 0 0 50]{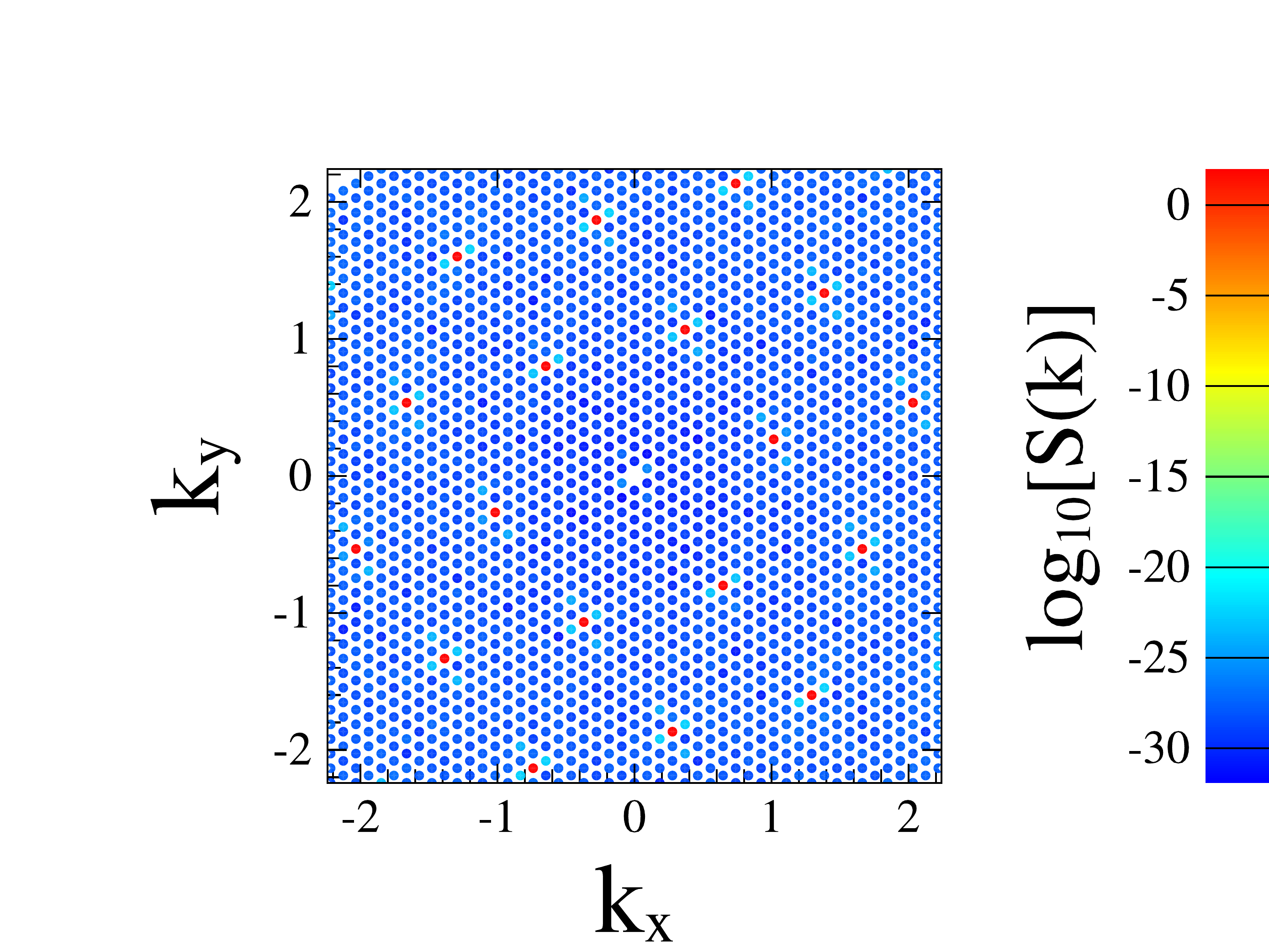}

\end{center}
\caption{
(Color online) Shown on the left are four representative numerically obtained ground-state configurations at $\chi=0.6818\ldots$ (first row), $\chi=0.7121\ldots$ (second row), $\chi=0.7424\ldots$ (third row), and $\chi=0.7878\ldots$ (fourth row). On the right are their corresponding structure factors, where colors indicate intensity values at reciprocal lattice points.
}
\label{fig_highchi}
\end{figure*}

Representative numerically obtained ground-state configurations and their structure factors (in logarithmic scales) are presented in Figs.~\ref{fig_lowchi}-\ref{fig_highchi}. For $0.5303\ldots \le \chi \le 0.6060\ldots$, the ground-state manifold appears to contain a variety of structures (see Fig.~\ref{fig_lowchi}). Except for the first one, all real-space configurations in Fig.~\ref{fig_lowchi} appear to be Bravais lattices. However, their structure factors are not as simple as a collection of Bragg peaks among a zero-intensity background, suggesting that the real-space configurations are not perfect Bravais lattices.

At $\chi=0.6363\ldots$ and $\chi=0.6666\ldots$ a type of relatively-simple-looking configuration appears (see Fig.~\ref{fig_medchi_tri}). The real-space configurations appear to be comprised of straight lines of particles with wavelike displacements relative to each other. The structure factors, on the other hand, consist of straight lines of nonzero values in a background of virtually zero ($<\ee{-20}$) intensities.

For $\chi \ge 0.6818\ldots$, the results are similar to that in Fig.~\ref{fig_medchi_tri}, but there exist so many constraints that the nonzero-value lines in the structure factor have to be interrupted. The interruptions grow in length as $\chi$ increases and eventually, at $\chi=0.7878\ldots$, the only nonzero structure factors are the Bragg peaks and the real-space configuration becomes a Bravais lattice.


\section{Analytical model of two-dimensional stacked-slider phase}
\label{model2}

\begin{figure}[H]
\begin{center}
\includegraphics[width=0.45\textwidth]{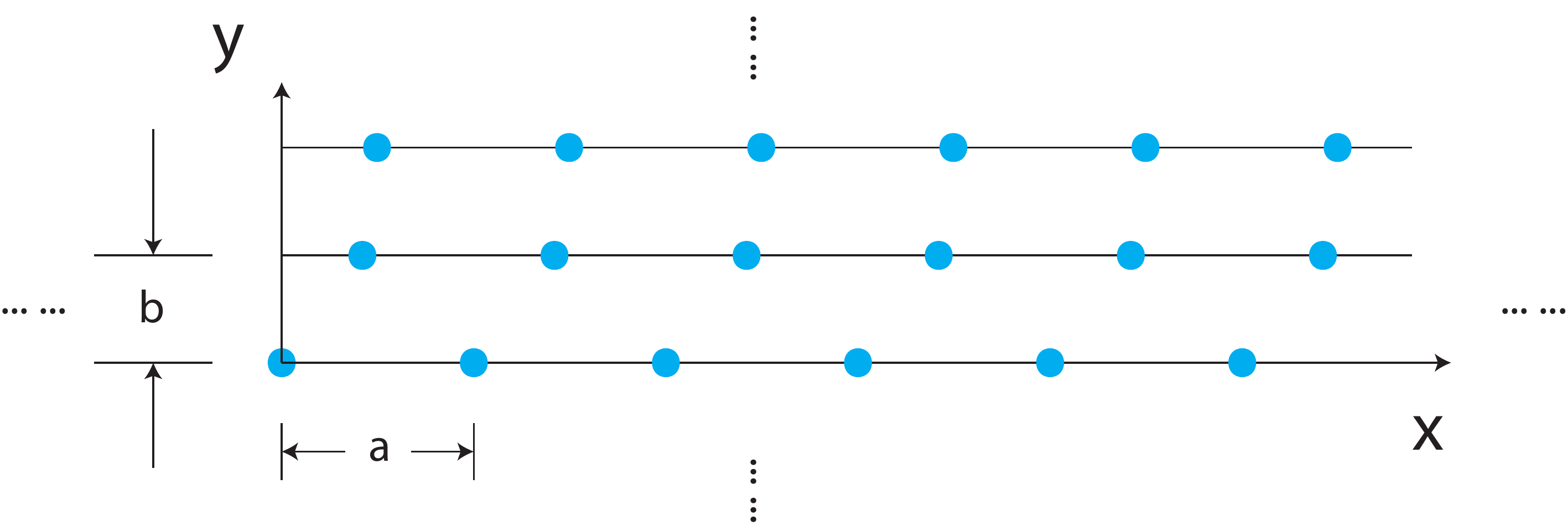}
\end{center}
\caption{(Color online) Schematic plot of the two-dimensional stacked-slider phase model. Each horizontal line of particles [indicated by large (blue) dots] form a one-dimensional integer lattice with lattice spacing $a$. Then multiple horizontal integer lattices are stacked vertically, with spacing $b$. Each horizontal line of particles can be translated freely to slide with respect to each other. }
\label{fig_model2}
\end{figure}

In this section we look closer at the simulation results that yield stacked-slider phases to see if an exact analytical construction can be extracted. We will see that understanding the configuration shown in Fig.~\ref{fig_medchi_tri} is the key to understanding other configurations. The real-space configuration in Fig.~\ref{fig_medchi_tri} seems to be made of straight horizontal lines that are displaced relative to each other.
Are the displacements of different horizontal lines independent of each other or correlated in some way? To answer this question, we numerically constructed a configuration that is made of horizontal straight lines of particles, just like the one shown in Fig.~\ref{fig_medchi_tri}, but with independent random displacements along each horizontal line. The structure factor of the new configuration has exactly the same support [the set of $\mathbf k$'s such that $S(\mathbf k) \neq 0$] as the one shown in Fig.~\ref{fig_medchi_tri}. Thus, the new configuration is also a ground state at this $\chi$ value. Therefore, the displacements of each line do not need to be correlated in any way. This allows us to find a two-dimensional stacked-slider phase model, depicted in Fig.~\ref{fig_model2}.

This analytical model allows the calculation of various properties of the two-dimensional stacked-slider phases.
One can find the analytical pair correlation function and structure factor of this model, assuming that the displacement of each line is independent and uniformly distributed between 0 and $a$. The pair correlation function $g_2(\mathbf r)$ is defined such that $\rho g_2(\mathbf r) d\mathbf r$ is the conditional probability that a particle is found in the volume element ${d\bf r}$ about ${\bf r}$, given that there is a particle at the origin. For the two-dimensional stacked-slider phase, $g_2(\mathbf r)$ can be found directly from the definition of this model:
\begin{equation}
g_2(x, y)=b \sum_{j \neq 0} \delta(y-bj) + ab \delta(y)\sum_{j \neq 0} \delta(x-aj),
\label{g2}
\end{equation}
where $x$ and $y$ are horizontal and vertical coordinates, both summations are over all nonzero integers $j$, and $\delta$ denotes the Dirac delta function. The structure factor $S(\mathbf k)$ can be found by Fourier transforming $g_2(\mathbf r)-1$:
\begin{equation}
S(\mathbf k) = 1+\rho F[g_2(\mathbf r)-1],
\label{sk}
\end{equation}
where $F[ \cdots ]$ denotes Fourier transform. Substituting \eqref{g2} into \eqref{sk}, one gets:
\begin{equation}
\begin{split}
S(\mathbf k) = \frac{2 \pi \delta(k_x)}{a} \left(\frac{2\pi}{b} \operatorname{III}_{2\pi/b}(k_y)-1 \right) \\ + \frac{2\pi}{a}\operatorname{III}_{2\pi/a}(k_x) - \frac{4\pi^2}{ab}\delta(k_x)\delta(k_y),
\end{split}
\label{sk2}
\end{equation}
where $k_x$ and $k_y$ denote the horizontal and vertical components of $\mathbf k$, respectively, and $\operatorname{III}_{T}(t) = \sum_{j=-\infty}^{+\infty} \delta(t-jT)$ is the Dirac comb function. Both the pair correlation function \eqref{g2} and the structure factor \eqref{sk2} are anisotropic, since swapping $x$ and $y$ in \eqref{g2} and \eqref{sk2} gives different expressions.

A topological property this model can predict is the connectedness of the ground-state manifold, i.e., whether or not a ground state can be continuously deformed to another ground state without crossing any energy barrier. Each stacked-slider configuration is obviously continuously connected to a rectangular lattice by the sliding motion of different lines. However, there are many permutations of the rectangular lattice. Are these permutations connected to each other through vertical and horizontal sliding motions? In the Appendix, we show that for a finite-size rectangular lattice consisting of $N$ particles, all permutations are connected if and only if $N$ is even.

Having found an analytical model of the ground states in this $\chi$ range, we move on to lower and higher $\chi$ ranges. The lower $\chi$ simulation results appear to be more complex. The first configuration in Fig.~\ref{fig_lowchi} appears to be similar to our existing analytical model, except that the nonzero-value regions in the structure factor are not strictly lines: The highest-intensity lines [$S(\mathbf k) \sim \ee{0}$] are surrounded by lower-intensity regions [$S(\mathbf k) \sim \ee{-10}$], which are surrounded by even lower-intensity regions [$S(\mathbf k) \sim \ee{-20}$]. The structure factor in the lower-intensity regions are very small, but are still much larger than the machine precision. (We use double-precision numbers, which have around 16 significant digits, to calculate ${\tilde n}({\bf k})$. Therefore, the machine precision of $S(k)=|{\tilde n}({\bf k})|^2/N$ should be on the order of $(\ee{-16})^2/N=\ee{-34}$.) So a natural question arises: Are the lower-intensity regions real or are they an artifact of finite-precision simulations?

\begin{figure}[h]
\begin{center}
\begin{tabular}{l c}
\multirow{1}{*}{(a)} & \raisebox{-.5\height}{\includegraphics[width=0.28\textwidth, trim=100 200 100 150]{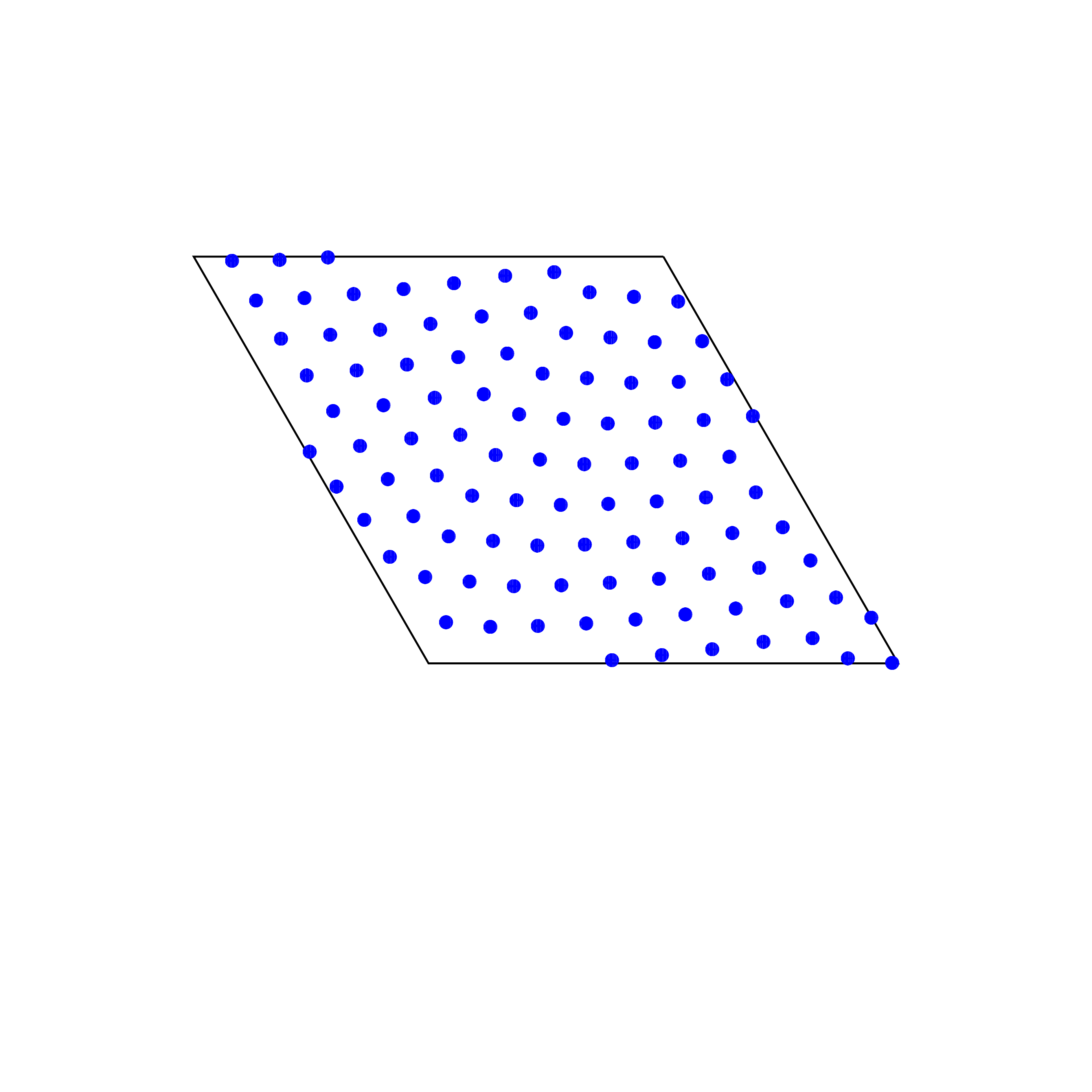}}\\
(b) & \raisebox{-.5\height}{\includegraphics[width=0.32\textwidth, trim=0 0 0 50]{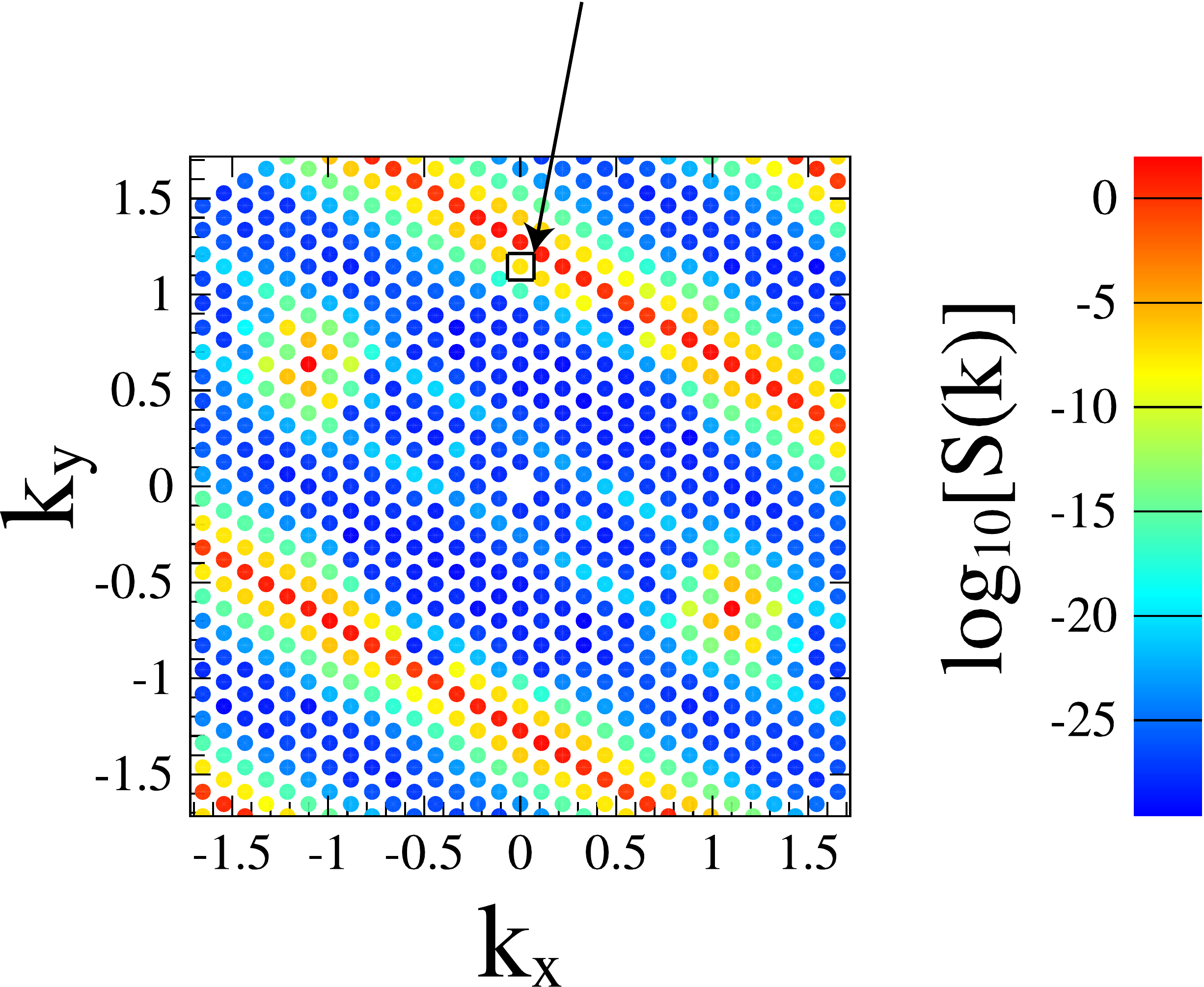}}\\
(c) & \raisebox{-.5\height}{\includegraphics[width=0.3\textwidth]{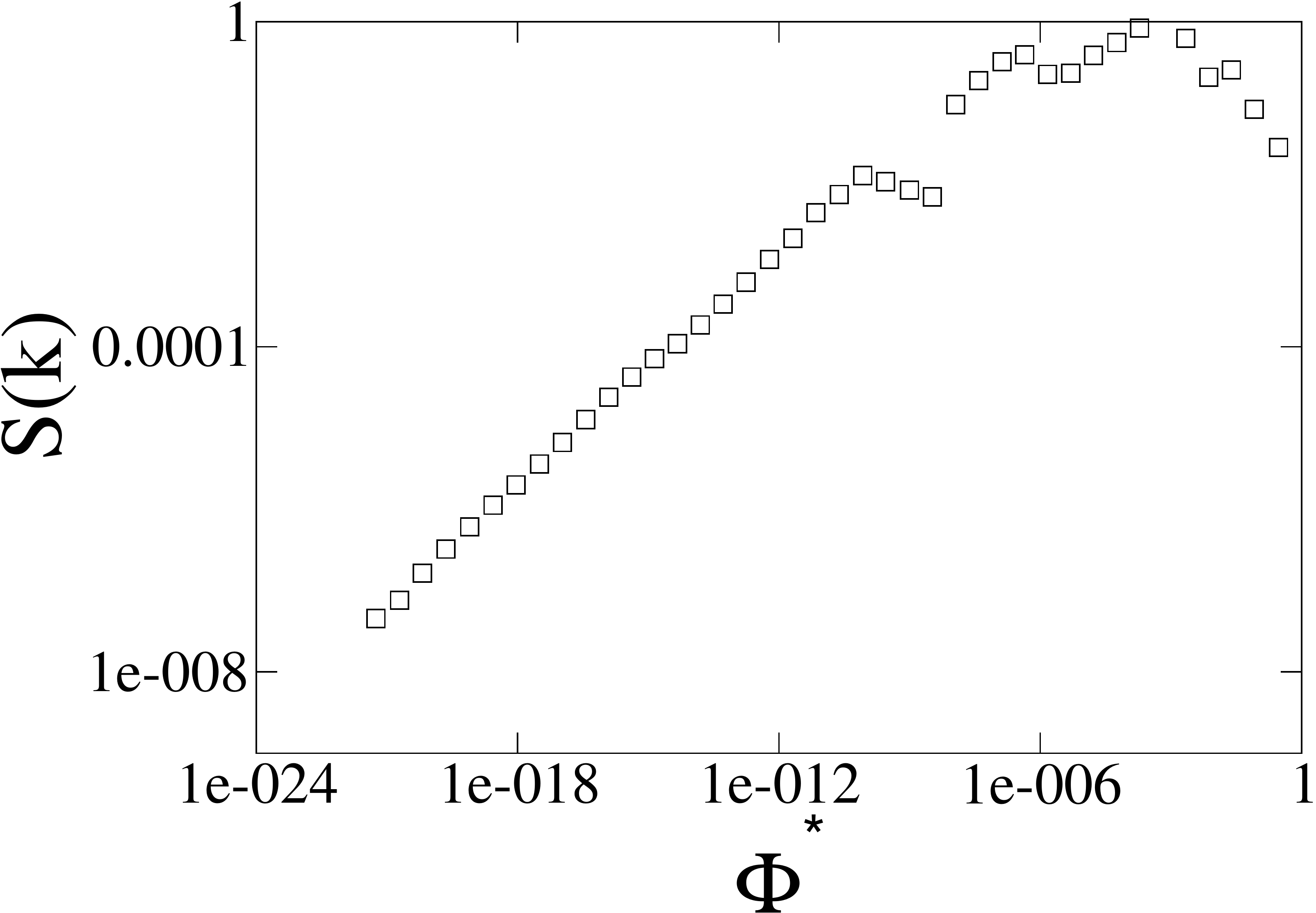}}\\
\end{tabular}
\end{center}
\caption{(Color online) (a) A numerically obtained ground state at $\chi=0.5606\ldots$. (b) The corresponding structure factor. A specific $\mathbf k$ point is indicated by a black square and an arrow. (c) The structure factor at this particular $\mathbf k$ point is plotted against total energy $\Phi^* $ during the optimization, showing $S(\mathbf k) \to 0$ as $\Phi^*  \to 0$.
}
\label{fig_lowchi_lin}
\end{figure}

To answer this question, we chose a $\mathbf k$ point right next to the highest-intensity line and plotted the structure factor at this $\mathbf k$ point versus the potential energy during the energy minimization (see Fig.~\ref{fig_lowchi_lin}). As $\Phi^* $ goes to zero, the structure factor at this $\mathbf k$ point also goes to zero. Thus, we believe the lower-intensity regions are the result of numerical imprecision. If one could carry out an infinite-precision simulation and drive this configuration to a true ground state, the structure factors in the lower-intensity regions should go to zero and the configuration would become consistent with our analytical model.

\begin{figure}[h]
\begin{center}
\begin{tabular}{l c}
(a) & \raisebox{-.5\height}{\includegraphics[width=0.28\textwidth, trim=100 200 100 150]{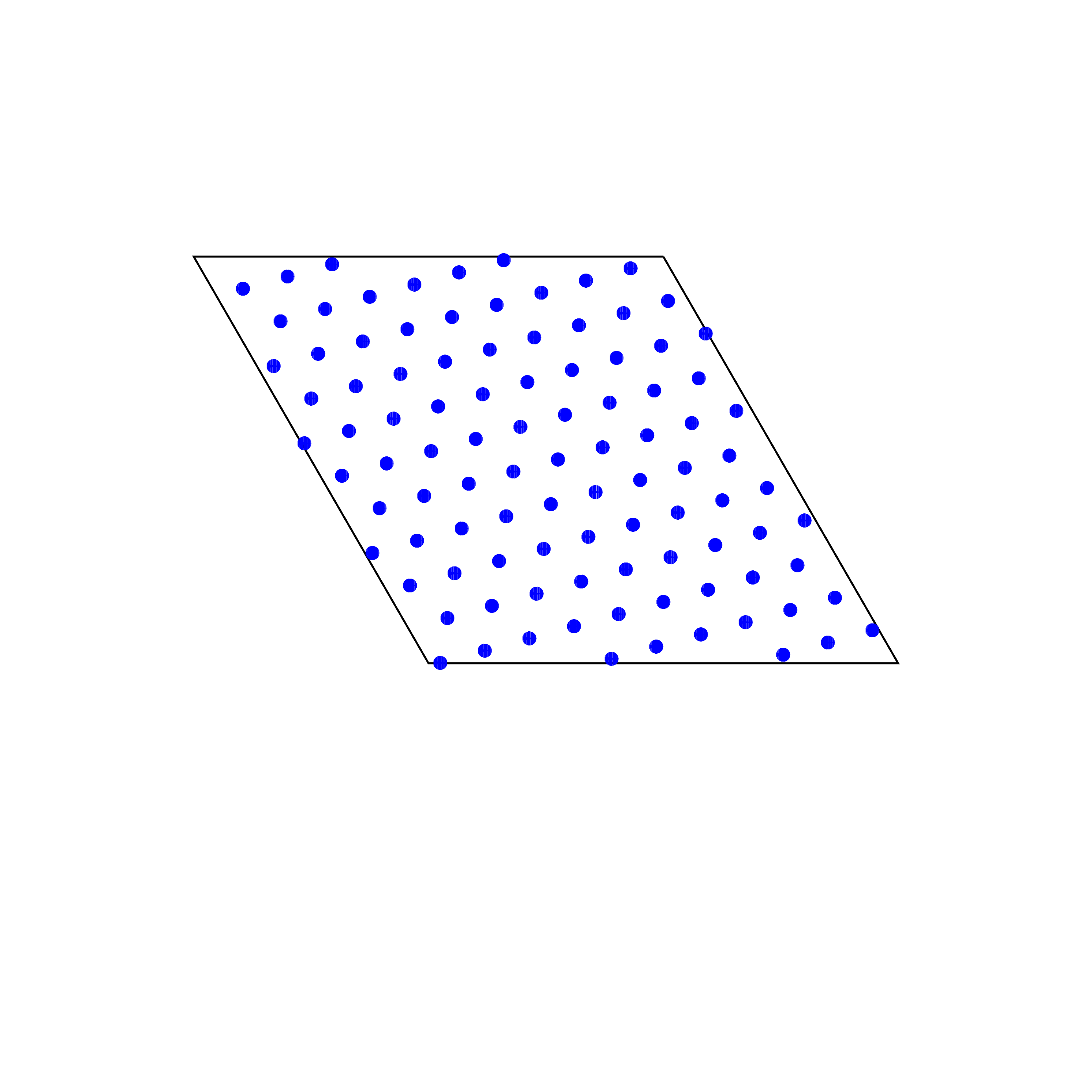}}\\
(b) & \raisebox{-.5\height}{\includegraphics[width=0.32\textwidth, trim=0 0 0 50]{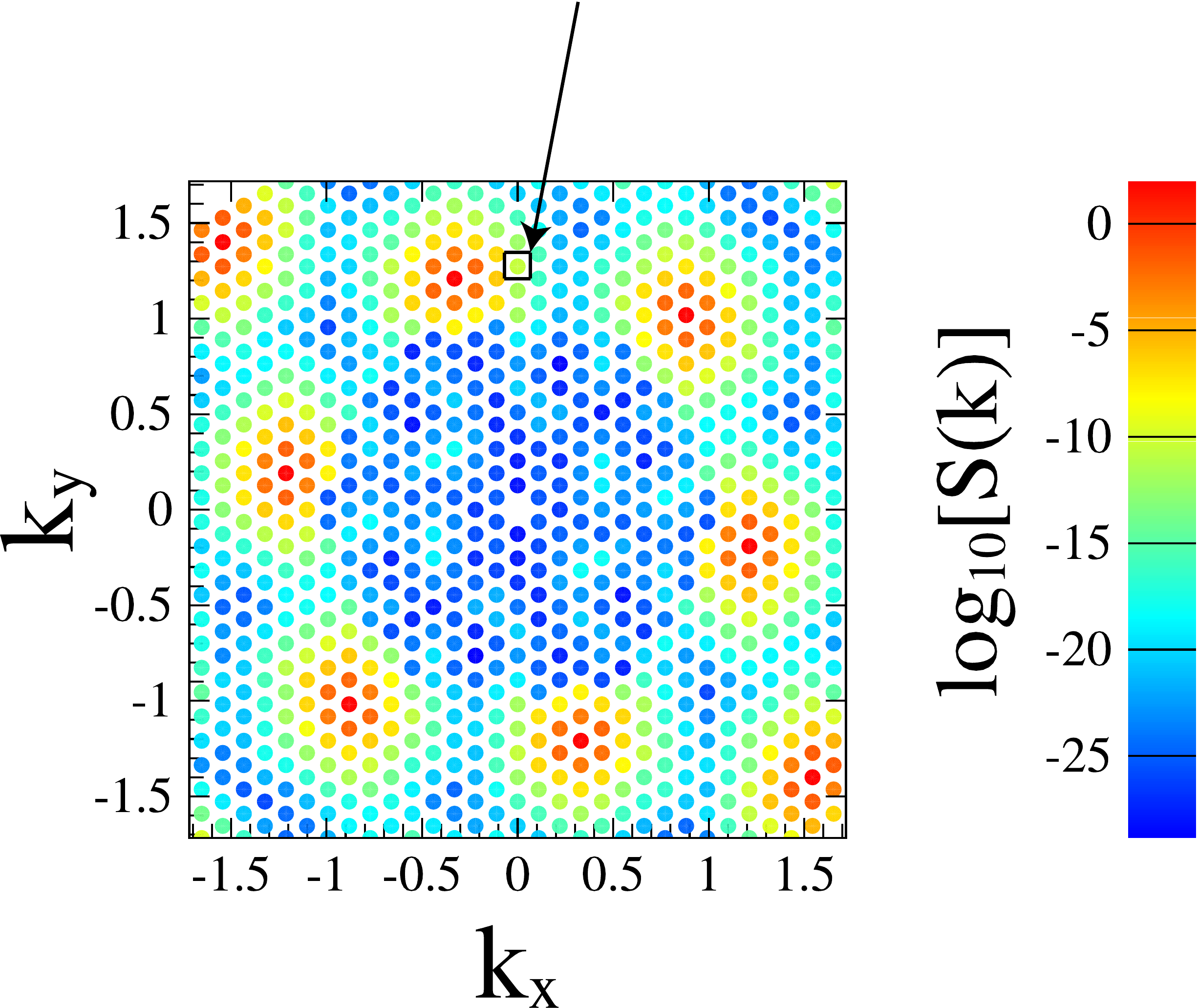}}\\
(c) & \raisebox{-.5\height}{\includegraphics[width=0.3\textwidth]{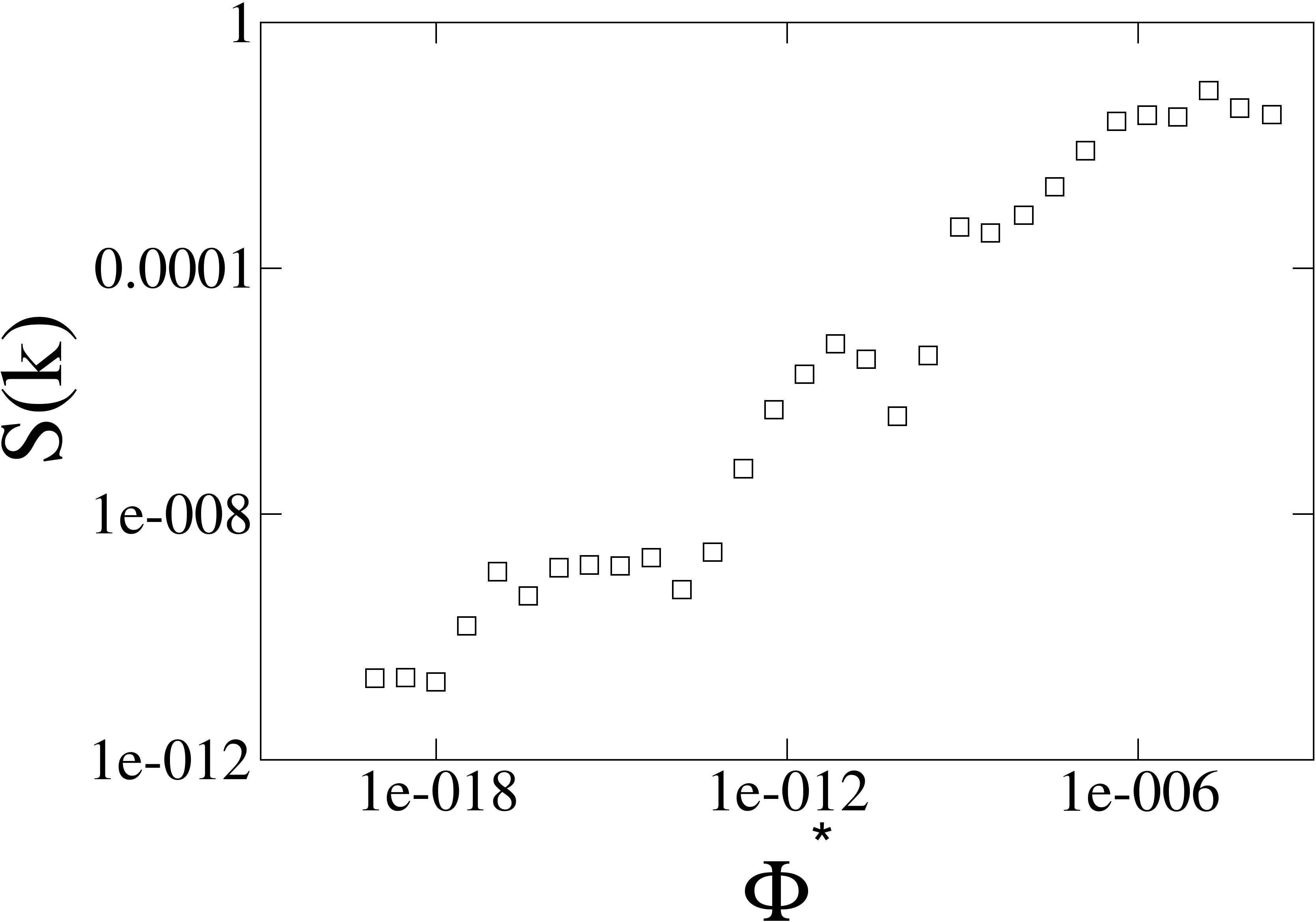}}\\
\end{tabular}
\end{center}
\caption{(Color online) (a) A numerically obtained ground state at $\chi=0.5606\ldots$. (b) The corresponding structure factor. A specific $\mathbf k$ point is indicated by a black square and an arrow. (c) The structure factor at this particular $\mathbf k$ point is plotted against total energy $\Phi^* $ during the optimization, showing $S(\mathbf k) \to 0$ as $\Phi^*  \to 0$.
}
\label{fig_lowchi_tri}
\end{figure}

Having understood the first configuration in Fig.~\ref{fig_lowchi}, let us move on to other configurations in that figure. The second and third configurations appear to be intermediate configurations between the first one and the fourth one. The fourth configuration looks like a Bravais lattice, except that the Bragg peaks are smeared out. Again, to find out whether this broadening of the Bragg peaks is real or artificial, we plotted the structure factor at a $\mathbf k$ point near a Bragg peak versus the potential energy in Fig.~\ref{fig_lowchi_tri}. We find again that the structure factor at this $\mathbf k$ point goes to zero as $\Phi^* $ goes to zero. Thus, the smearing out of the Bragg peaks is also due to numerical imprecision. If one could carry out an infinite-precision energy minimization on this configuration, one should get a Bravais lattice.

So far we have demonstrated that the numerically obtained ground states follow a simple model at $\chi=0.6363\ldots$ and $\chi=0.6666\ldots$. We have also demonstrated that while the numerically obtained ground states for $0.5303\ldots \le \chi < 0.6363\ldots$ appear to be richer, they are actually exactly the same as either the model or a Bravais lattice if we could perform infinite-precision simulations. 
However, as we move to higher $\chi$'s, the ground states start to lose degrees of freedom.
As shown in Fig.~\ref{fig_highchi}, at $\chi=0.6818\ldots$, the high-intensity lines in the structure factor develop zero-intensity interruptions. In our stacked-slider phase model, if each line of particles could move independently, then the high-intensity lines in the structure factor would have no interruptions. Thus, these interruptions indicate constraints in the displacements of each line of particles. At $\chi=0.7121\ldots$, the lines are interrupted even further, indicating even more constraints in the displacements of each line. At $\chi=0.7424\ldots$, the structure becomes a two-particle-basis crystal. Eventually, at $\chi=0.7878\ldots$, the structure becomes a Bravais lattice.

Starting from $\chi = 0.6818\ldots$, the stacked-slider phase become more constrained as $\chi$ increases. To study how constrained this phase is at different $\chi$ values, we calculate the number of zero eigenvalues $n_e$ of the Hessian matrix of the potential energy. This number is equal to the number of independent ways to deform the structure such that the energy scales more slowly than quadratic, which is an upper bound of the dimensionality of the ground-state configuration space $n_c$ [i.e. the number of independent ways to deform the structure such that the $\Phi^* ({\bf r}^N)$ remains zero]. For $\chi < 0.6818\ldots$, our model predicts $n_c=11$ (since there are two translational degrees of freedom, and nine independent ways to slide the ten lines of particles relative to each other) and our calculation also find $n_e=11$. At $\chi = 0.6818\ldots$, $0.7121\ldots$, $0.7424\ldots$, and $0.7878\ldots$, our calculations find $n_e=9$, $5$, $3$, and $2$, respectively. This calculation suggests that as $\chi$ increases, $n_c$ gradually decreases. Eventually, $n_c=2$, indicating that there is no way to deform the structure other than trivial translations.

\section{Generalized Stacked-slider phase model}
\label{model}

We now generalize the two-dimensional stacked-slider phase model to higher dimensions. To begin with, we present and prove the following theorem:

{\sl Stealthy Stacking Theorem. } Let $d_P$ and $d_Q$ be positive integers. Let $W$ be $(d_P+d_Q)$-dimensional Euclidean space.
Let $W_P$ be a $d_P$-dimensional subspace of $W$ and $W_Q$ be the $d_Q$-dimensional orthogonal complement space of $W_P$.
Let $P$ be a point pattern in $W_P$ with density $\rho_P$. For each point $\mathbf a \in P$, let $Q(\mathbf a)$ be a point pattern in $W_Q$ with some density $\rho_Q$ independent of $\mathbf a$.
If $P$ is stealthy up to certain reciprocal-space cutoff $K_P$ and all $Q(\mathbf a)$'s are stealthy up to certain reciprocal-space cutoff $K_Q$ in their subspace, then the following point pattern in $W$,
\begin{equation}
\{\mathbf a + \mathbf b | \mathbf a \in P, \mathbf b \in Q(\mathbf a)\}
\label{Overlay_Pattern}
\end{equation}
is a stealthy point pattern up to $K=min(K_P, K_Q)$.

\vspace{0.3in}

{\sl Proof.} The collective density variable of the point pattern in Eq.~(\ref{Overlay_Pattern}) is
\begin{equation}
\displaystyle {\tilde n}(\mathbf k) = \sum_{\mathbf a \in P} \sum_{\mathbf b \in Q(\mathbf a)} \exp [-i \mathbf k \cdot (\mathbf a + \mathbf b)].
\end{equation}
Since $W_P$ and $W_Q$ are two orthogonal complementary subspaces of $W$, we can divide vector $\mathbf k$ into two parts $\mathbf k = \mathbf k_P+\mathbf k_Q$, where $\mathbf k_P \in W_P$ and $\mathbf k_Q \in W_Q$. Therefore,
\begin{equation}
\begin{split}
\displaystyle {\tilde n}(\mathbf k) = \sum_{\mathbf a \in P} \sum_{\mathbf b \in Q(\mathbf a)} \exp [-i (\mathbf k_P+\mathbf k_Q) \cdot (\mathbf a + \mathbf b)] \\
= \sum_{\mathbf a \in P} \exp (i \mathbf k_P \cdot \mathbf a ) \sum_{\mathbf b \in Q(\mathbf a)} \exp (-i \mathbf k_Q \cdot \mathbf b).
\label{c3}
\end{split}
\end{equation}
For any $\mathbf k$ such that $0<|\mathbf k|\le K$, $|\mathbf k_Q| \le |\mathbf k|\le K \le K_Q$. If $\mathbf k_Q \neq \mathbf 0$, then the stealthiness of point patterns $Q(\mathbf a)$ gives
\begin{equation}
\sum_{\mathbf b \in Q(\mathbf a)} \exp (-i \mathbf k_Q \cdot \mathbf b) =0
\end{equation}
and therefore $\displaystyle {\tilde n}(\mathbf k)=0$. On the other hand, if $\mathbf k_Q = \mathbf 0$, then $\mathbf k_P=\mathbf k$ and Eq.~(\ref{c3}) becomes
\begin{equation}
\displaystyle {\tilde n}(\mathbf k) = N_{Q(\mathbf a)} \sum_{\mathbf a \in P} \exp (-i \mathbf k \cdot \mathbf a ),
\end{equation}
where $N_{Q(\mathbf a)}$ is the number of particles in pattern $Q(\mathbf a)$, which is independent of $\mathbf a$ because all the $Q(\mathbf a)$'s have the same density.
Since $0<|\mathbf k|\le K \le K_P$, the stealthiness of point pattern $P$ gives
\begin{equation}
\displaystyle {\tilde n}(\mathbf k) =0.
\end{equation}
To summarize, for any $\mathbf k$ such that $0<|\mathbf k|\le K$, whether or not $\mathbf k_Q = \mathbf 0$, ${\tilde n}(\mathbf k)$ is always zero. Therefore, the point pattern (\ref{Overlay_Pattern}) is stealthy up to $K$.

\begin{figure}[h]
\begin{center}
\includegraphics[width=0.45\textwidth]{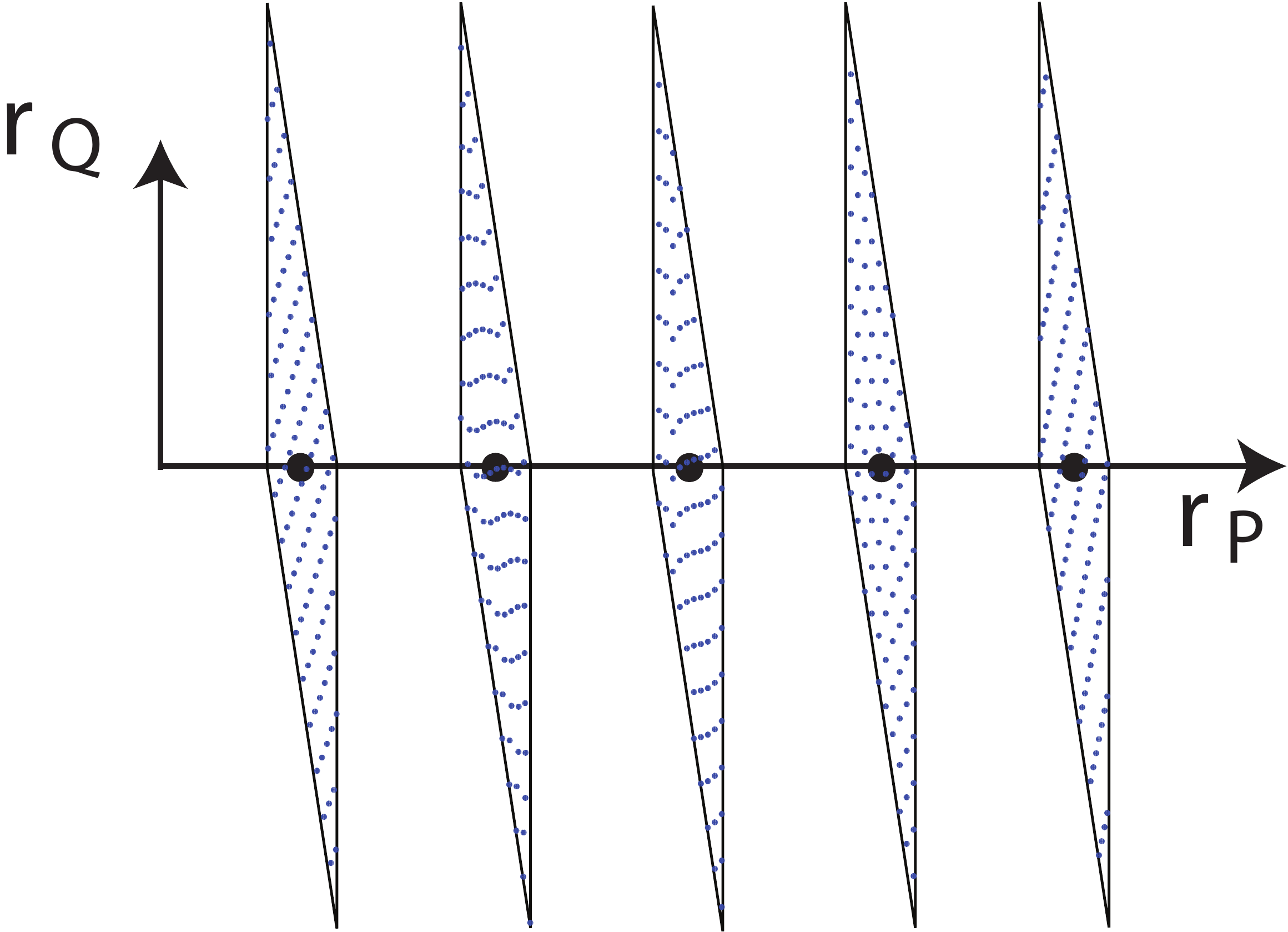}
\end{center}
\caption{(Color online) Schematic plot of the stacked-slider phase model. The large black dots form an integer lattice (point pattern $P$). By replacing each black dot with a two-dimensional stealthy point pattern (indicated by small blue dots) of the same density [point patterns $Q(\mathbf a)$], the overall three-dimensional point pattern consisting of all the small blue dots is stealthy. The two vectors $\mathbf r_P$ and $\mathbf r_Q$ are in subspaces $W_P$ and $W_Q$, respectively. Note that since some $Q(\mathbf a)$'s are two-dimensional stacked-slider configurations, this configuration allows both interlayer and intralayer sliding motions, as detailed in Sec.~\ref{model2}.
}
\label{fig_model}
\end{figure}

The parameter $\chi$ of this point pattern can be calculated using Eq.~(35) of Ref.~\cite{torquato2015ensemble}. Our calculation yields
\begin{equation}
\chi=\frac{v_1(d_P+d_Q; K)}{2 (d_P+d_Q) (2\pi)^{d_P+d_Q} \rho_P \rho_Q},
\label{overlay_chi}
\end{equation}
where $v_1(d; r)$ is the volume of a $d$-dimensional hypersphere of radius $r$.
In the case $K_P=K_Q$, using Eq.~(35) of Ref.~\cite{torquato2015ensemble}, Eq.~(\ref{overlay_chi}) can be simplified to:
\begin{equation}
\chi=\frac{2 v_1(d_P+d_Q; 1)}{v_1(d_P; 1)v_1(d_Q; 1)} \frac{d_P d_Q}{d_P+d_Q} \chi_P \chi_Q.
\label{overlay_chi_2}
\end{equation}

\begin{figure}
\begin{center}
\includegraphics[width=0.45\textwidth]{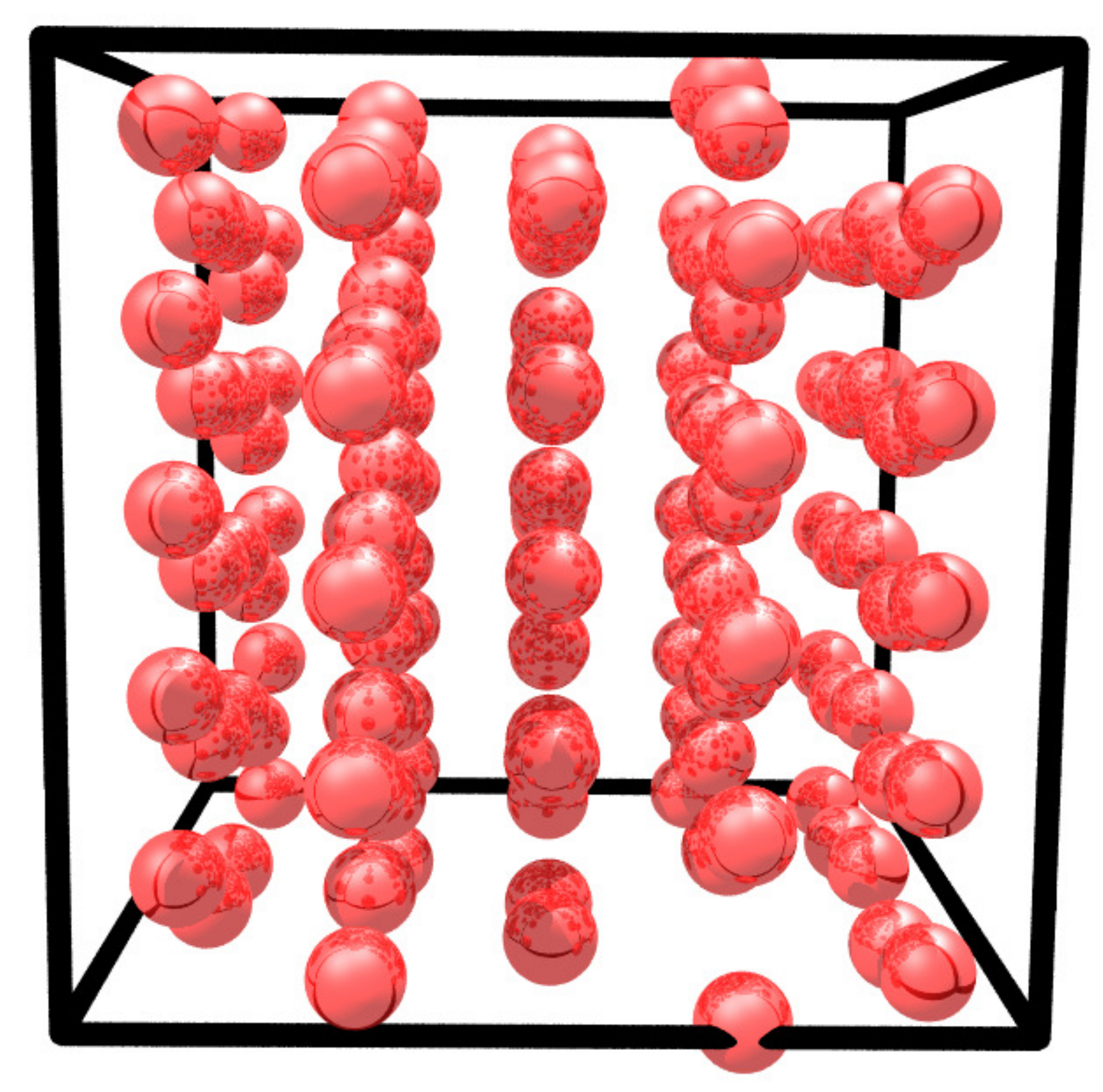}
\end{center}
\caption{(Color online) Three-dimensional stacked-slider configuration stealthy up to $\chi=0.6981\ldots$. This configuration is obtained by sliding each vertical plane of particles relative to each other and then sliding each vertical line in each plane relative to each other starting from the simple cubic lattice.
}
\label{fig_WavySimpleCubic}
\end{figure}

The aforementioned theorem allows us to construct stacked-slider configurations in higher dimensions. To construct a stacked-slider configuration in $d \ge 2$, choose two lower dimensions $d_P$ and $d_Q$ such that $d_P+d_Q=d$. Choose a $d_P$-dimensional stealthy configuration $P$ and replace each particle $\mathbf a$ in $P$ with a $d_Q$-dimensional stealthy configuration $Q(\mathbf a)$ and the resulting $d$-dimensional configuration is a stacked-slider one. The resulting configuration is often anisotropic, since $d_P$ dimensions are treated separately from the remaining $d_Q$ dimensions. See Fig.~\ref{fig_model} for an illustration of a three-dimensional stacked-slider configuration with $d_P=1$ and $d_Q=2$. 

Certain three-dimensional crystal structures can allow sliding deformations while remaining stealthy at relatively large (greater than $0.5$) $\chi$. As Fig.~\ref{fig_WavySimpleCubic} shows, the simple cubic lattice allows the sliding motion of each two-dimensional square-lattice layer and the sliding motion of each line of particles inside every layer for $\chi$ up to $0.6981\ldots$. Barlow packings \cite{sloane1998kepler}, including the face-centered-cubic packing and the hexagonal close packing, also allow the sliding motion of each triangular-lattice layer of particles for $\chi$ up to $0.7600\ldots$.

Equation~\eqref{overlay_chi_2} can be used to calculate the maximum $\chi$ values of the stacked-slider-phase, $\chi_{max}^{ss}$, assuming unconstrained sliding motions, in each space dimension $d$. To do this one can try all possible combinations of positive integers $d_P$ and $d_Q$ such that $d_P+d_Q=d$, and let $\chi_P$ and $\chi_Q$ equal to $\chi_{max}^*$ in $d_P$ and $d_Q$ dimensions, respectively. Our calculations for $2\le d\le 4$ are summarized in Table~\ref{ss_chi}. There is no obvious trend in these low dimensions. However, as $d$ increases, the factor $\frac{2 v_1(d_P+d_Q; 1)}{v_1(d_P; 1)v_1(d_Q; 1)} \frac{d_P d_Q}{d_P+d_Q}$ in Eq.~\eqref{overlay_chi_2} decreases for any $d_P$ and $d_Q$. Thus, $\chi_{max}^{ss}$ should become arbitrarily small in sufficiently high dimensions.

\begin{table}[h]
\caption{Comparison of the maximum $\chi$ value of stacked-slider-phases predicted by the generalized model $\chi_{max}^{ss}$ and the maximum $\chi$ value of Bravais lattices $\chi_{max}^*$ in two, three, and four dimensions.}
\begin{tabular}{|c | c| c |c|}
\hline
$d$ &$\chi_{max}^{ss}$&$ \chi_{max}^* $&$ \chi_{max}^{ss}/\chi_{max}^*$ \\
\hline
2 &$ \pi/4 $&$ \pi/\sqrt{12} $& 0.8660...\\\hline
3 &$ \frac{4\pi}{9\sqrt{3}} $&$ \frac{2\sqrt{2}\pi}{9} $& 0.8712...\\\hline
4 &$ \frac{\sqrt{2}\pi^2}{16} $&$\frac{\pi^2}{8} $& 0.7071...\\
\hline
\end{tabular}
\label{ss_chi}
\end{table}

Similar to two-dimensional stacked-slider configurations, the higher-dimensional ones also have implicit constraints [i.e. $\mathbf k$ vectors such that $|\mathbf k| > K$ and $S(\mathbf k)=0$]. As seen in Eq.~\eqref{c3}, $S(\mathbf k)=|{\tilde n}(\mathbf k)|^2/N=0$ as long as $0 < |\mathbf k_Q| \le K$. One can thus choose arbitrarily large $\mathbf k_P$ such that $|\mathbf k| = |\mathbf k_P+\mathbf k_Q| > K$.

\section{Feasible region of the configuration space}
\label{region}

Although stacked-slider configurations are part of the ground-state manifold of stealthy potentials, we will show in this section that they are not entropically favored, as indicated in Ref.~\onlinecite{torquato2015ensemble}. Entropically favored ground states are the configurations that most likely appear in the canonical ensemble in the zero-temperature limit \cite{torquato2015ensemble}. In this limit, as a good approximation, the system can only visit part of the configuration space where $\Phi^* ({\bf r}^N)$ [in Eq.~\eqref{pot_f3}] is less than $\epsilon$, where $\epsilon>0$ tends to zero as the temperature tends to zero. This part of the configuration space is therefore called the feasible region. If the feasible region corresponding to one set of the ground states is much smaller than the entire feasible region in the configuration space, this set will almost never appear in the canonical ensemble, i.e., they are not entropically favored.

In the infinite-system-size limit, the feasible region of any stacked-slider configuration
is much smaller than that of any crystal if both the stacked-slider configuration and the crystal
are ground states.
This is because as $N \to \infty$, the configurational dimension $n_c$ [i.e., the number of independent ways to deform the structure such that the $\Phi^* ({\bf r}^N)$ remains zero] of stacked-slider phases scales more slowly than the number of particles $N$. For example, for a two-dimensional stacked-slider configuration in which each row of particles can slide independently, $n_c$ scales as $\sqrt{N}$. As discussed in Sec.~\ref{model2}, the number of zero eigenvalues of the Hessian matrix of the potential energy $n_e$ is equal to $n_c$. Since a nonzero eigenvalue of the Hessian matrix corresponds to a quadratic scaling in one direction, in the $dN$-dimensional configuration space, $\Phi^* ({\bf r}^N)$ has quadratic scaling in $dN-n_c$ directions. In these directions, as $\epsilon \to 0$, the width of the feasible region scales as $\sqrt{\epsilon}$. In the remaining $n_c$ directions, the width of the feasible region is much larger, since these directions correspond to translations of different rows of particles, which keeps $\Phi^* ({\bf r}^N)$ zero. If we let the widths of the feasible region in these directions be $L$, then the total volume of the feasible region of the stacked-sliding phase is approximately
\begin{equation}
V_s \approx L^{n_c} \epsilon^{(dN-n_c)/2} \approx L^{\sqrt{N}} \epsilon^{(dN-\sqrt{N})/2}.
\end{equation}

\begin{figure}[H]
\begin{center}
\includegraphics[width=0.45\textwidth]{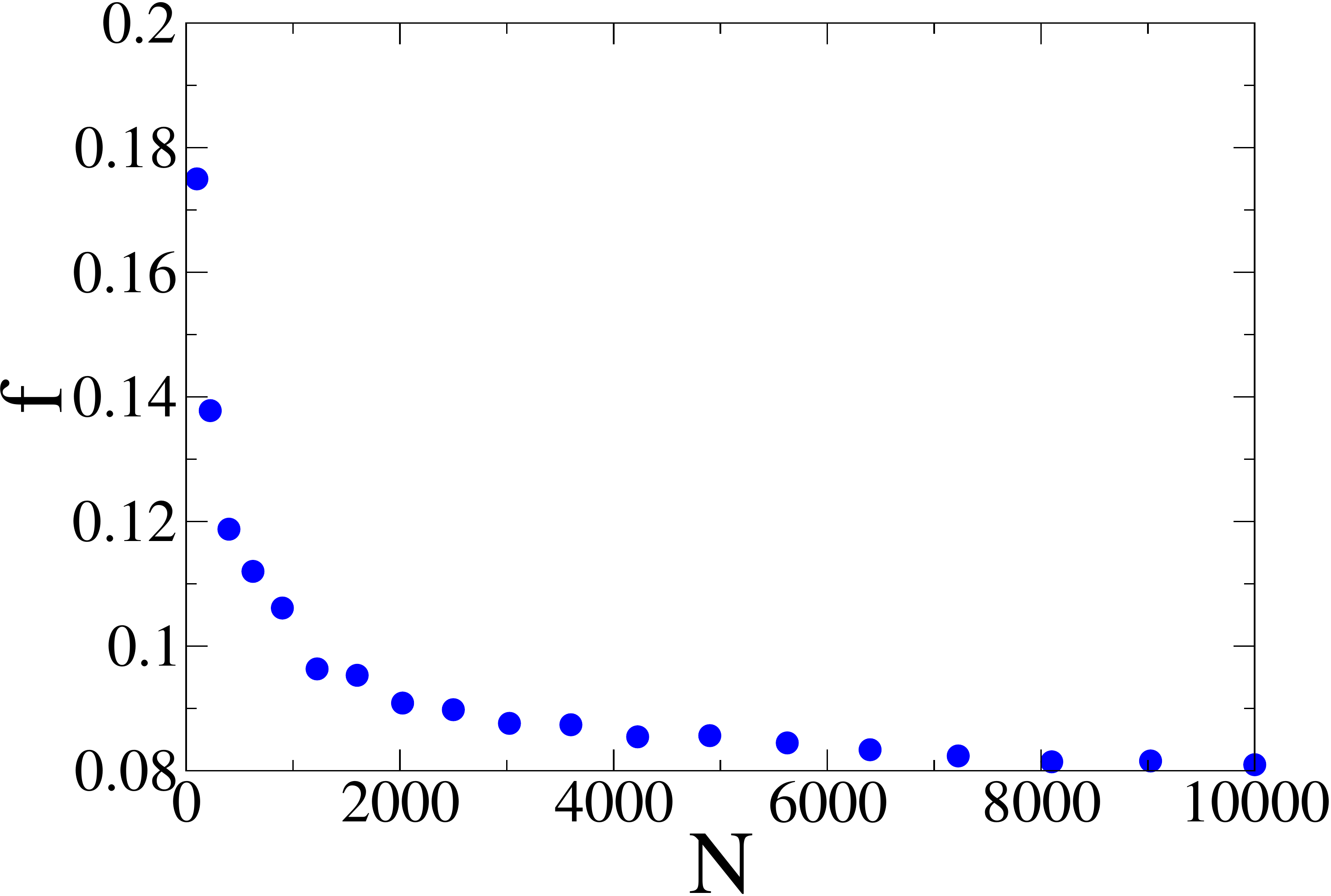}
\end{center}
\caption{(Color online) Fraction of zero eigenvalues of the Hessian matrix of the potential energy $\displaystyle f=\frac{n_e}{dN}$ for triangular lattices of various numbers of particles $N$ at $\chi=0.6$.}
\label{fig_eigen}
\end{figure}

In the case of a crystalline structure, $n_e$ scales as $N$ when $N \to \infty$. This can be seen in Fig.~\ref{fig_eigen}, where we plot $\displaystyle f={n_e}/dN$ versus $N$ for triangular lattices at $\chi=0.6$. This figure shows that $f$ tends to some constant as $N$ grows, which means $n_e$ scales as $N$. Since a zero eigenvalue of the Hessian matrix of $\Phi^* ({\bf r}^N)$ implies a slower-than-quadratic scaling in some direction, the width of the feasible region in these $n_e$ directions scales larger than $\sqrt{\epsilon}$ as $\epsilon \to 0$. Let the widths of the feasible region in these $n_e$ directions be $\epsilon^x$, where $0<x<1/2$ is some exponent. The width of the feasible region in the remaining $dN-n_e$ directions scales as $\sqrt{\epsilon}$. The total volume of the feasible region of a crystal is approximately
\begin{equation}
V_c \approx \epsilon^{(dN-n_e)/2}\epsilon^{n_ex} \approx \epsilon^{dN(1-f)/2}\epsilon^{dNfx}.
\end{equation}
The ratio of $V_s$ and $V_c$ is approximately
\begin{equation}
\frac{V_s}{V_c} \approx L^{\sqrt{N}} \epsilon^{[dNf(1-2x)-\sqrt{N}]/2}.
\end{equation}
Since $x<1/2$, as $N \to \infty$ and $\epsilon \to 0$, $\frac{V_s}{V_c} \to 0$. Therefore, the feasible region of the stacked-slider phase is much smaller than that of the crystal. Since there are always crystalline structures competing with the stacked-slider phase, the latter is never entropically favored.

\section{relative stability of stacked-slider phases}
\label{stability}
We have shown that the feasible region of stacked-slider phases is always smaller than that of crystal phases and thus concluded that stacked-slider phases are never equilibrium phases at $T=0$. 
This conclusion is confirmed by low-temperature molecular dynamics simulations reported in Ref.~\onlinecite{zhang2015ground}, which found disordered structures for $\chi<1/2$ and crystalline structures for $\chi>1/2$. However, this simple conclusion cannot explain or predict energy minimization results from high-temperature initial configurations that were used previously \cite{uche2004constraints}, where a transition from disordered phases to metastable stacked-slider phases was observed as $\chi$ increases, characterized by the change of the support of $S(\mathbf k)$. In two dimensions, Ref.~\onlinecite{uche2004constraints} reported that this transition is at $\chi=0.57\ldots$, but high-fidelity simulations, reported in Sec.~\ref{numerical}, produced stacked-slider configurations at $\chi=0.5305\ldots$, suggesting that the transition is earlier than $0.5305\ldots$. Another observation on the disordered region supports our result: Section V of Ref.~\onlinecite{batten2009interactions} reported that the fraction of normal modes with vanishing frequency $f$ in disordered phases is exactly $1-2\chi$ for $\chi<1/2$. However, this exact relation cannot be true for the $\chi>1/2$ region, since $f$ is non-negative. This suggests that there exists a sharp transition at $\chi=1/2$, which is likely the phase transition to the stacked-slider phase. Although Ref.~\onlinecite{batten2009interactions} only reported the relation $f=1-2\chi$ in two dimensions, it explained this relation by simple counting arguments involving the number of constraints versus the number of degrees of freedom and hence this relation should apply in any dimension. Therefore, for any $d$, as long as stacked-slider phases exist for some $\chi$ above 1/2, there should be a {\it nonequilibrium} phase transition from disordered phases to stacked-slider phases at the threshold $\chi=1/2$.

It is noteworthy that one dimension is an exception of the above discussion.
Previously, the existence of implicit constraints [$\mathbf k$'s such that $|\mathbf k| > K$ and $S(\mathbf k)=0$] was often used to distinguish stacked-slider phases from disordered phases \cite{uche2004constraints, batten2009interactions}. Therefore, one-dimensional stealthy ground states in the range $1/3<\chi<1/2$, proven to have implicit constraints \cite{fan1991constraints}, were considered to be stacked-slider phases \cite{batten2009interactions}. However, this study suggests that one-dimensional stealthy ground states in this range are not a typical stacked-slider phase. First, our model only predicts stacked-slider phases if the space dimension $d$ is a sum of two positive integers $d=d_P+d_Q$. This requires that $d \ge 2$. Second, the $\chi$ range of the one-dimensional stealthy ground states with implicit constraints is also very different from that of the higher-dimensional stacked-slider phases. We also found that one-dimensional stealthy ground states in this $\chi$ range satisfy the relation $f=1-2\chi$ and can be obtained from energy minimizations starting from random initial configurations with $100\%$ success rate; both are characteristics of disordered phases \cite{batten2009interactions}.

\section{Conclusions and discussion}
\label{conclusion}

\begin{table*}
\caption{Comparison of the properties of some common states of matter. Here crystals and quasicrystals signify perfect crystals and perfect quasicrystals, respectively, without any defects (e.g., phonons and phasons). The checks and crosses indicate whether or not different phases have the attributes listed in the first column. }
\renewcommand{\arraystretch}{2}
\begin{tabular}{|c || c| c |c| c| c| c|}
\hline
Property & \parbox{1.5cm}{Crystals \\ \cite{sands2012introduction}} & \parbox{2cm}{Quasicrystals \\ \cite{shechtman1984metallic, levine1984quasicrystals, levine1986quasicrystals, bindi2009natural, dotera2014mosaic}}& \parbox{2cm}{Stacked-slider \\ phases} & \parbox{3cm}{Disordered \\ ground states of \\ stealthy potentials \\ \cite{uche2004constraints, uche2006collective, batten2008classical, batten2009interactions, torquato2015ensemble}}& \parbox{2.2cm}{Liquid crystals \\ \cite{chandrasekhar1992liquid}} & \parbox{1.3cm}{Liquids \\ \cite{tabor1991gases}}\\
\hline
periodicity & \cmark & \xmark & \xmark & \xmark & \xmark & \xmark \\\hline
\parbox{3cm}{positive \\ shear modulus} & \cmark & \cmark & \xmark & \xmark & \xmark & \xmark \\\hline
hyperuniformity & \cmark & \cmark & \cmark & \cmark & \xmark & \xmark \\\hline
anisotropy & \cmark & \cmark & \cmark & \xmark & \cmark & \xmark \\\hline
\parbox{3cm}{long-range  orientational order} & \cmark & \cmark & \cmark & \xmark & \cmark & \xmark \\\hline

\end{tabular}
\label{phases}
\end{table*}

In this paper we studied using numerical and theoretical techniques stacked-slider phases, which are metastable states that are  part of the ground-state manifold of stealthy potentials at densities in which crystal ground states
are favored entropically in the canonical
ensemble in the zero-temperature limit \cite{torquato2015ensemble, zhang2015ground}. 
The numerical results suggested analytical models of this phase in two, three and higher dimensions. Utilizing this model, we estimated the size of the feasible region of the stacked-slider phase, finding it to be smaller than that of crystal structures in the infinite-system-size limit, which
is consistent with our recent previous work \cite{torquato2015ensemble, zhang2015ground}. 
In two dimensions, we also determined exact expressions for the pair correlation function and structure factor of the analytical model of stacked-slider phases, and analyzed the connectedness of the ground-state manifold of stealthy potentials in this density regime.

Our analytical constructions demonstrate that stacked-slider phases are nonperiodic, statistically anisotropic structures
that possess long-range orientational order but have zero shear modulus. 
Since stacked-slider phases are part of the ground-state manifold of stealthy potentials, they are also hyperuniform. 
Therefore, stacked-slider phases are
distinguishable states of matter that are uniquely different from some
common states of matter listed
in Table~\ref{phases}.
Note that distinctions between the attributes indicated in the table
may be subtly different.
For example, crystals, quasicrystals, and stacked-slider phases all have long-range orientational order, but with different symmetries. While crystals can only have twofold, threefold, fourfold, or sixfold rotational symmetries, quasicrystals have prohibited crystallographic rotational symmetries.
Stacked-slider phases generally do not have any rotational symmetry, but the fact that they can be constructed by stacking lower-dimensional stealthy configurations in a higher-dimensional space makes the stacking directions different from the sliding directions, giving them their unique orientational order.

Our understanding of stacked-slider phases is only in its infancy with many open questions. For example, what
is the nature of the associated excited
states? Can stacked-slider phases emerge from particles interacting with other potentials not necessarily as ground states? Can such phases 
be entropically favored in some ensemble and with what other phases would it coexist? This is just a partial list of possible of future avenues of research in our understanding of this unusual phase of matter.

\begin{acknowledgments}
This research was supported by the U.S. Department of Energy, Office of Basic Energy Sciences, Division of Materials Sciences and Engineering under Grant No. DE-FG02-04-ER46108.
\end{acknowledgments}

\appendix
\section{Connectedness of permutations of 2D stacked-slider phase}
\label{connectedness}

As discussed in Sec.~\ref{model2}, each two-dimensional stacked-slider configuration is connected to a permutation of the rectangular lattice. Therefore, a natural question is whether or not these permutations of the rectangular lattice are also connected through sliding motions. If all permutations of the rectangular lattice are connected, then the entire stacked-slider phase ground-state manifold is connected. We will show that, for a rectangular lattice consisting of $A$ rows and $B$ columns of particles, if each row and each column can slide individually, then all permutations of the rectangular lattice are connected if and only if $AB$ is even. We will number all the particles from 1 to $AB$. Each permutation will be represented by an $A \times B$ matrix. Three different sliding motions will be frequently used in this section. They are as follows:
\begin{itemize}
\item Move the top row of particles to the right by one particle spacing, denoted by $\stackrel{\rightarrow}{\Rightarrow}$;
\item move the leftmost column of particles upward by one particle spacing, denoted by $\stackrel{\uparrow}{\Rightarrow}$;
\item and move the leftmost column of particles downward by one particle spacing, denoted by $\stackrel{\downarrow}{\Rightarrow}$.
\end{itemize}

As an example of this notation, for $A=B=2$, permutations $\begin{pmatrix} 1 & 2 \\ 3 & 4 \end{pmatrix}$ and $\begin{pmatrix} 2 & 1 \\ 3 & 4 \end{pmatrix}$ are connected because
\begin{equation}
\begin{pmatrix} 1 & 2 \\ 3 & 4 \end{pmatrix} \stackrel{\rightarrow}{\Rightarrow} \begin{pmatrix} 2 & 1 \\ 3 & 4 \end{pmatrix}.
\label{p21}
\end{equation}
Similarly, permutations $\begin{pmatrix} 1 & 2 \\ 3 & 4 \end{pmatrix}$ and $\begin{pmatrix} 3 & 2 \\ 1 & 4 \end{pmatrix}$ are connected because
\begin{equation}
\begin{pmatrix} 1 & 2 \\ 3 & 4 \end{pmatrix} \stackrel{\downarrow}{\Rightarrow} \begin{pmatrix} 3 & 2 \\ 1 & 4 \end{pmatrix}.
\label{p22}
\end{equation}
So far we have demonstrated that it is possible to swap the two adjacent particles in the first row [by Eq.~\eqref{p21}] or the two adjacent particles in the first column [by Eq.~\eqref{p22}] for $A=B=2$. Since the system has translational symmetry, one can swap any two adjacent particles. The swapping of any two nonadjacent particles can be done by a series of adjacent-particle swapping. For example, to swap nonadjacent particles 1 and 4 in $\begin{pmatrix} 1 & 2 \\ 3 & 4 \end{pmatrix}$, one can swap particles 1 and 2, then swap particles 1 and 4, and then swap particles 2 and 4. Finally, since we can swap any two particles, we can connect one permutation to any other permutation by swapping each particle with the particle in its new place. Therefore, all permutations of $2 \times 2$ rectangular lattices are connected by row-sliding and column-sliding movements.

Next, we show that one can swap two adjacent particles for $A=3$ and $B=4$. To swap the first two particles in the first row, one can perform the following sliding operations:
\begin{multline}
\begin{pmatrix}1&2&3&4\\5&6&7&8\\9&10&11&12\end{pmatrix}\stackrel{\rightarrow}{\Rightarrow}
\begin{pmatrix}4&1&2&3\\5&6&7&8\\9&10&11&12\end{pmatrix}\stackrel{\uparrow}{\Rightarrow}
\begin{pmatrix}5&1&2&3\\9&6&7&8\\4&10&11&12\end{pmatrix}\\\stackrel{\rightarrow}{\Rightarrow}
\begin{pmatrix}3&5&1&2\\9&6&7&8\\4&10&11&12\end{pmatrix}\stackrel{\downarrow}{\Rightarrow}
\begin{pmatrix}4&5&1&2\\3&6&7&8\\9&10&11&12\end{pmatrix}\stackrel{\rightarrow}{\Rightarrow}
\begin{pmatrix}2&4&5&1\\3&6&7&8\\9&10&11&12\end{pmatrix}\\\stackrel{\uparrow}{\Rightarrow}
\begin{pmatrix}3&4&5&1\\9&6&7&8\\2&10&11&12\end{pmatrix}\stackrel{\rightarrow}{\Rightarrow}
\begin{pmatrix}1&3&4&5\\9&6&7&8\\2&10&11&12\end{pmatrix}\stackrel{\rightarrow}{\Rightarrow}
\begin{pmatrix}5&1&3&4\\9&6&7&8\\2&10&11&12\end{pmatrix}\\\stackrel{\downarrow}{\Rightarrow}
\begin{pmatrix}2&1&3&4\\5&6&7&8\\9&10&11&12\end{pmatrix}.
\label{p41}
\end{multline}
To swap the first two particles in the first column, one can perform the following sliding operations starting from the third-to-last configuration in Eq.~\eqref{p41}:
\begin{equation}
\begin{pmatrix}1&3&4&5\\9&6&7&8\\2&10&11&12\end{pmatrix}\stackrel{\downarrow}{\Rightarrow}
\begin{pmatrix}2&3&4&5\\1&6&7&8\\9&10&11&12\end{pmatrix}\stackrel{\rightarrow}{\Rightarrow}
\begin{pmatrix}5&2&3&4\\1&6&7&8\\9&10&11&12\end{pmatrix}.
\label{p42}
\end{equation}

Equations~\eqref{p41}~and~\eqref{p42} shows the steps to swap the first two particles in the first row, or the first two particles in the first column, for $A=3$ and $B=4$. This can be generalized to any $A>3$ and any even $B>4$. The generalization to $A>3$ is more obvious because the same steps can be directly applied to any $A$ and achieve the same goal. The generalization to larger even $B$ is less obvious. For this case, one needs to repeat the first four operations in Eq.~\eqref{p41} $(B/2-1)$ times and then perform the rest of the steps in Eq.~\eqref{p41} or ~\eqref{p42}.
Since it is possible to swap any two adjacent particles for any $A$ and any even $B$ , from the same argument as the $A=B=2$ case, all permutations of particles for any $A$ and any even $B$ are also connected. Similarly, all permutations of particles for any even $A$ and any $B$ are also connected because a $90^\circ$ rotation turns it to the even $B$ case. Therefore, all permutations are connected as long as $AB$ is even.

When $AB$ is odd, not all permutations are connected. This is because none of the sliding operations change the parity of the permutation. Thus, two permutations with different parity cannot be connected with any combinations of sliding operations.


\begin{thebibliography}{30}
\bibitem{successrate}The reader will notice that the success rate reported in Table~\ref{Nt} is dramatically nonuniform. This is another unusual feature of this class of phases that deserves further attention.
\end{thebibliography}
\end{document}